\documentstyle[12pt,times]{article}
\newlength{\abstractwidth}
\renewcommand{\thefootnote}{\fnsymbol{footnote}}
\renewcommand{\thanks}[1]{\footnote{#1}} 
\newcommand{\starttext}{
\setcounter{footnote}{0}
\renewcommand{\thefootnote}{\arabic{footnote}}}

\newcommand{\be}{\begin{equation}}
\newcommand{\bea}{\begin{eqnarray}}
\newcommand{\eea}{\end{eqnarray}}
\newcommand{\beq}{\begin{equation}}
\newcommand{\ee}{\end{equation}}
\newcommand{\eeq}{\end{equation}}
\newcommand{\N}{{\cal N}}
\newcommand{\<}{\langle}
\newcommand{\ophi}{{\cal O}_\phi}
\newcommand{\oc}{{\cal O}_C}
\renewcommand{\>}{\rangle}
\def\ba{\begin{eqnarray}}
\def\ea{\end{eqnarray}}
\newcommand{\PSbox}[3]{\mbox{\rule{0in}{#3}\includegraphics{#1}\hspace{#2}}}
\renewcommand{\S}[1]{\Sigma_{#1}}
\newcommand{\hd}{{\Delta\over 2}}
\newcommand{\D}[1]{\Delta_{#1}}
\newcommand{\G}{\Gamma}
\newcommand{\g}[1]{\Gamma\left(#1\right)}
\newcommand{\half}{1\over 2}

\def\tr{{\rm tr}}
\def\12{{1 \over 2}}
\def\32{{3 \over 2}}
\def\72{{7 \over 2}}
\def\92{{9 \over 2}}
\def\d{{d \over 2}}

\def\phat{\hat {\cal O}_\phi}
\def\chat{\hat {\cal O}_C}
\def\ob{{\cal O}_B}
\def\N{{\cal N}}
\def\O{{\cal O}}

\def\tr{{\rm tr}}
\def\half{{1 \over 2}}
\def\d{{d \over 2}}

\begin{document}
\begin{titlepage}
\bigskip
\hskip 3.7in\vbox{\baselineskip12pt
\hbox{UCLA/99/TEP/46}
\hbox{SU-ITP-99-51}
\hbox{hep-th/99mmnnn}}
\bigskip\bigskip\bigskip\bigskip

\centerline{\Large \bf The Operator Product Expansion of
\boldmath ${\cal N}=4$  SYM}
\medskip
\centerline{\Large \bf and the 4--point Functions of Supergravity}

\bigskip\bigskip
\bigskip\bigskip

\centerline{   \bf Eric D'Hoker$^{\,a}$,
Samir D. Mathur$^{\,b}$, }
\medskip
\centerline{  \bf Alec Matusis$^{\,c}$,
Leonardo Rastelli$^{\,d,}$\footnote[1]{\tt e-mails:
dhoker@physics.ucla.edu, mathur@pacific.mps.ohio-state.edu,
 alecm@leland.Stanford.edu, rastelli@ctp.mit.edu.} }
\bigskip
\bigskip
\centerline{$^a$ \it Department of Physics}
\centerline{ \it University of California, Los Angeles, CA 90095}
\smallskip
\centerline{$^b$ \it Department of Physics}
\centerline{  \it Ohio State University,  Columbus { OH}  43210}
\smallskip
\centerline{$^c$ \it Department of Physics}
\centerline{ \it  Stanford University, Stanford, {CA} 94305}
 \smallskip
\centerline{$^d$ \it Center for Theoretical Physics}
\centerline{ \it Massachusetts Institute of Technology, 
Cambridge, { MA} 02139}
\bigskip\bigskip
\begin{abstract}
We give a detailed Operator Product Expansion interpretation
of the results for  conformal 4--point functions
computed from supergravity through the AdS/CFT duality.
We show that for an arbitrary scalar exchange in $AdS_{d+1}$ all the
power--singular terms
in the direct channel limit (and only these terms) exactly match the
corresponding contributions to the OPE of the operator
dual to the exchanged bulk field and of its conformal descendents.
The leading
logarithmic singularities in the 4--point functions of  protected
${\cal N}=4$
super--Yang
Mills operators (computed from IIB supergravity on $AdS_5 \times S^5$)
 are interpreted as $O(\frac{1}{N^2})$ renormalization
effects of
the double--trace products appearing in the OPE.
Applied to
the   4--point functions
of the operators
 $\ophi \sim ({\rm tr}F^2 + \dots)$
and  $\oc \sim ({\rm tr}F \tilde F + \dots)$, this analysis
leads to the prediction  that the double--trace composites $[\ophi \oc]$ and
 $[\ophi \ophi -\oc \oc]$ have anomalous dimension
$-\frac{16}{N^2}$ in the large $N$,
 large $g_{YM}^2N$ limit. We describe a geometric picture of the
OPE in the dual gravitational theory,
for both the power--singular terms and the leading
logarithms. We comment on several possible extensions 
of our results.

\end{abstract}

\end{titlepage}
\starttext
\baselineskip=18pt
\setcounter{footnote}{0}

\section{Introduction}

The study of 4--dimensional Conformal Field Theories
is an old and important topic.
The AdS/CFT correspondence  \cite{maldacena,review} 
provides new powerful tools
to address this problem. Difficult dynamical questions
about the strong coupling  behavior of the CFT are answered by
perturbative computations in Anti de Sitter supergravity.
A natural set of questions
concerns the nature of the Operator Product Expansion
of the CFT at strong coupling. Thanks to the AdS/CFT duality, we can
now answer some of these questions.

The prime example of an exactly conformal 4--dimensional
field theory, namely the
${\cal N}=4$ Super--Yang Mills theory with gauge group $SU(N)$,
is dual \cite{maldacena}  to Type IIB string theory on $AdS_5 \times S^5$,
with $N$ units of 5--form flux and compactification radius
$R^2 = \alpha'  (g_{YM}^2 N)^{\frac{1}{2}}$.
 For large $N$ and large 't Hooft coupling
$\lambda \equiv g_{YM}^2 N$ the dual string theory is approximated by
weakly coupled supergravity in $AdS_5$ background.
Since the 5--dimensional Newton constant $G_5 \sim R^3/N^2$, the
perturbative expansion in supergravity
corresponds to the $1/N$ expansion in the CFT.
Correlation functions of local operators of the CFT
belonging to short multiplets of the superconformal algebra
are given for $N \rightarrow \infty$, $\lambda \rightarrow \infty$
by supergravity amplitudes according to the prescription
of \cite{polyakov,witten}. While the supergravity results for
2-- and 3--point functions
of chiral operators \cite{april}--\cite{chalmers1}
have been found to agree with the
free field approximation, giving strong evidence for the existence of
non--renormalization theorems \cite{dhfskiba}--\cite{nonrenorm},
 4--point functions \cite{talk}--\cite{other4points}
 certainly contain some non--trivial
dynamical information\footnote{Perturbative studies of 4--point functions
in ${\cal N}=4$ SYM include \cite{kovacs}--\cite{howewest4pt}.}.

The 4--point
functions of the operators
 $\ophi$ and $\oc$ dual to the dilaton and axion
fields
were obtained
in \cite{march99} through a supergravity computation,
and  expressed as very explicit  series expansions
in terms of two conformal invariant variables. The fundamental fields of the
${\cal N}=4$ theory are the gauge boson $A_\mu$, 4 Majorana fermions
$\lambda^a$ and 6 real scalars $X^i$, all in the adjoint representation
of the gauge group $SU(N)$. The operators $\ophi$ and $\oc$
are the exactly marginal operators that correspond to changing
the gauge coupling and the theta angle of the theory. In other terms
the SYM Lagrangian has the form ${\cal L} \sim \frac{1}{g_{YM}^2}\,\ophi
+ \frac{\theta}{8 \pi^2} \,\oc$. 
It is convenient to define operators
that have unit--normalized 2--point functions,
$\hat \ophi \sim \frac{1}{N}\, \ophi \sim \frac{1}{N}\, {\rm tr} (F^2 +\dots)$ and
 $\hat \oc \sim \frac{1}{N}\, \oc \sim \frac{1}{N}\,
{\rm tr} (F \tilde F +\dots)$.

The computation of the axion--dilaton 4--point functions
required the sum of several supergravity diagrams, weighted by
the appropriate couplings
in the Type IIB action on   $AdS_5 \times S^5$,
\be \label{action}
S = \frac{1}{2 \kappa_5^2} \int_{AdS_5} d^5z \,\sqrt{g} \;\left(
-{\cal R} +\Lambda +
\frac{1}{2}\,g^{\mu \nu} \partial_\mu \phi \partial_\nu \phi
+\frac{1}{2}\,e^{2 \phi} \,g^{\mu \nu} \partial_\mu C \partial_\nu C
\right).
\ee

Besides these `complete'  dilaton--axion 4--point functions,
explicit results for arbitrary supergravity
exchange diagrams involving massive scalars, massive vectors
and massless gravitons are also available \cite{howto}.
In the present paper we give
a detailed OPE interpretation of some of these
results, and obtain new  predictions for the strong coupling behavior
of the ${\cal N}=4$ SYM theory.
The fact that the 5d supergravity amplitudes can be consistently interpreted
in terms of a 4d local OPE is by itself quite remarkable,
and constitutes a strong test of the AdS/CFT duality.

Let us introduce the main issues from the field theory viewpoint.
By considering the limit of
a 4--point function as the operator locations become pairwise close
 (take a `t--channel' limit
$x_{13} \equiv |\vec x_1 -\vec x_3| \rightarrow 0$ and
$x_{24} \equiv |\vec x_2 -\vec x_4| \rightarrow 0$), we expect
a double OPE expansion to hold:
\be \label{ope}
 \< {\cal O}_1(\vec x_1) {\cal O}_2(\vec x_2) {\cal O}_3(\vec x_3)
{\cal O}_4(\vec x_4)
\> =\\
 \sum_{n,m}\,\frac{\alpha_{n}
\;\< {\cal O}_n(\vec x_1){\cal O}_m(\vec x_2) \>
\;\beta_m }{(x_{13})^{\Delta_1+\Delta_3-\Delta_m}(x_{24})^{\Delta_2+\Delta_4-\Delta_n}} \, ,
\ee
at least as an asymptotic series, and hopefully with a finite
radius of convergence in $x_{13}$ and $x_{24}$. For simplicity
we have suppressed all Lorentz and flavor structures and generically
denoted by $\{ {\cal O}_n \}$ the set of primary operators ${\cal O}_p$
and their conformal descendents $\bigtriangledown^k {\cal O}_p$.
Let us take the operators   in the 4--point function to be
`single--trace'\footnote{The trace is over the color group
$SU(N)$. Single--trace operators in SYM are dual to single--particle
Kaluza--Klein states
in supergravity \cite{extremal}.}
 chiral primaries ${\rm tr} (X^{(i_1} \dots X^{i_k)})$
or any of their superconformal descendents, such as
$\ophi \sim {\rm tr} (F^2 +\dots)$. These operators belong to short
representations of the superconformal algebra\footnote{We will 
call `chiral' any operator belonging to a short multiplet.}
and their dimensions
do not receive quantum corrections. On purely field--theoretic
grounds, we expect that all the operators
allowed by selection rules (most of which are not chiral)
can contribute
as intermediate states to the r.h.s. of (\ref{ope}).

The AdS/CFT duality makes the interesting prediction
that the non--chiral operators of the SYM theory actually fall into two
classes\footnote{Group theoretic aspects of string and multi--particle
states are considered in \cite{group}.}, which behave very differently at strong coupling:
\begin{itemize}
\item
Operators dual to string states, like for example the
Konishi operator ${\rm tr} X^{i} X^{i}$, whose dimensions
become very large in the strong coupling limit
(as $\sim \lambda^{\frac{1}{4}}$);
\item
Multi--trace operators obtained by taking (suitably regulated)
products 
of single--trace chiral operators at the same point, 
like for example the normal--ordered product
$[\ophi \ophi]$. These operators are dual to multi--particle
supergravity states.
\end{itemize}
Since in the limit of large $N$, large $\lambda$
the dual supergravity description is weakly coupled,
the dimension of a multi--particle state
 is approximately the sum of the dimensions of the
single--particle (single--trace in SYM language) constituents,  with
 small  corrections of order $G_5 \sim 1/N^2$ due to gravitational
interactions. From a perturbative analysis in the SYM
theory it is not hard to show (see Section \ref{fieldtheory}) that for
large $N$ and small $\lambda$ the anomalous
dimension of an operator like $[\ophi \ophi]$ is of the form
$ \sim \frac{1}{N^2} f(\lambda) +O(\frac{1}{N^4})$,
where  $f(\lambda)$ can be computed
as a perturbative series $f(\lambda) = \sum_{k \geq 1}
 a_k \lambda^k$. The AdS/CFT
duality then predicts that as $\lambda \rightarrow \infty$
the function $f(\lambda)$ saturates to a finite value.

A non--trivial issue
is whether
in the double OPE intepretation of  the supergravity amplitudes
one can find any remnant of the non--chiral
operators corresponding to string states. In fact, although these operators
acquire a large anomalous dimension as $\lambda
\rightarrow \infty$,
an infinite number of them is exchanged in
the r.h.s. of (\ref{ope}) for any finite $\lambda$,
and one may  worry about
a possible non--uniformity of the limit.
On the contrary, our analysis will
 lend support to the idea that as $\lambda \rightarrow \infty$
 the string states consistently decouple.

Since each single--trace chiral operator ${\cal O}$
of the SYM theory is dual to some
Kaluza--Klein mode $\phi$ of supergravity, there appears to be a
1--1 correspondence
between supergravity diagrams in which $\phi$ is exchanged in
the `t--channel' (that is, the bulk--to--bulk $\phi$
propagators joins the pairs ${\cal O}_1 {\cal O}_3$
and ${\cal O}_2 {\cal O}_4$, see Fig.1)
and the contribution
to the double OPE (\ref{ope}) of the operator
$\cal O$ and its conformal descendents
 $\bigtriangledown^k {\cal O}$. In Section 2 we prove a general
theorem\footnote{A similar result has been obtained in \cite{liu}
in a rather different formalism.}:
for any scalar exchange\footnote{We restrict for simplicity
to
pairwise equal external dimensions $\Delta_1 =\Delta_3 \leq \Delta_2 =\Delta_4
$.}
 in $AdS_{d+1}$
all the singular terms $O(\frac{1}{x_{13}^n})$
 (and only these terms)
exactly match the corresponding contributions of the conformal
block $\{ \bigtriangledown^k {\cal O} \}$ to the double OPE
(\ref{ope}) of the $d$--dimensional boundary CFT.
We believe that a similar theorem must hold for exchanges
of arbitrary spin.
The correspondence between supergravity exchanges and `conformal
partial waves' breaks down precisely when the double--trace
operator $[{\cal O}_1 {\cal O}_3]$, which has dimension
$\Delta_1 + \Delta_3 + O(1/N^2)$, starts contributing to the OPE.
This result implies that as $\lambda \rightarrow
\infty$ the singular part of the OPE
of two chiral SYM operators is entirely given by other
chiral operators and their multi--trace products.
This is of course consistent with the expectation
that non--chiral operators corresponding to string states have
a large dimension in this limit.

A generic feature of supergravity 4--point amplitudes is
that their asymptotic expansions contain logarithmic terms.
For example, a `t--channel' exchange diagram
(Fig.1) contains as $x_{13} \rightarrow 0$ a logarithmic singularity
of the form $\frac{1}{x_{12}^{2\Delta_1 +2\Delta_2}}
\log\left(\frac{x_{13}x_{24}}{x_{12}x_{34}}\right)$, as well as a whole series
of regular terms $x_{13}^n \log(x_{13})$.
(All these terms are subleading with respect to the
power singularities $O(\frac{1}{x_{13}^k})$ discussed above.
The expansion of the same diagram in the limit $x_{12} \rightarrow 0$
contains instead no power singularities,
and the logarithmic term is leading.)
This logarithmic behavior may appear puzzling in a unitary CFT.
However, as stressed to us by Witten  early in this work, logs naturally
arise in the perturbative expansion of a CFT 
from anomalous dimensions and operator mixing.

Section 3 of the paper
is devoted to a general discussion of the logarithmic behavior of
CFT's. The difference between logs that arise in a pertubative
expansion of a unitary theory and the `intrinsic logs' of a non--unitary
theory is emphasized. In our case the perturbative
parameter is
$1/N$. As already noted, we expect operators
like $[\ophi \ophi]$ to have anomalous dimensions of order
$O(\frac{1}{N^2})$. The logs in the supergravity 4--point functions
arise indeed at the correct order and with the right structure
to be interpreted as $O(\frac{1}{N^2})$ renormalization effects
of the double--trace composites produced in the OPE of two chiral
operators.

In Section 4, we perform a careful analysis of the leading
logarithmic terms in the supergravity correlators
$\< \ophi \ophi \ophi \ophi \>$ and $\<\ophi  \oc \ophi \oc \>$.
In order to reproduce the structure of the  supergravity logs it is crucial
to take into account the mixing between operators
with the same quantum numbers,
like $[\ophi \ophi]$ and $[\oc \oc]$. This analysis leads to the
prediction of the strong
coupling values of the anomalous dimensions of the operators
$[\ophi \oc]$ and $[\ophi \ophi -\oc \oc]$, which  are the only
two operators with the maximal
$U_Y(1)$ charge $|Y|=4$ and thus cannot mix with any other operator
of approximate dimension 8.

In Section 5 we present our conclusions and propose some avenues
for future research.

\setcounter{equation}{0}
\section{Supergravity Exchanges versus OPE: Power Singularities}

\begin{figure}
\begin{center} \PSbox{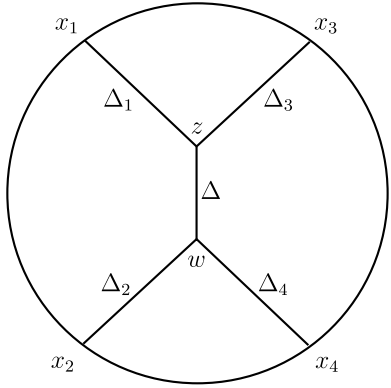 hscale=80 vscale=80}{7.1in}{5.4in} 
\vspace{-8cm}
\end{center}
\caption{Scalar exchange in the `t--channel'.}
\end{figure}
There is an intriguing relation \cite{talk} \cite{liutseytlin2}
between supergravity exchange
diagrams and `conformal partial waves'. A conformal partial wave
is the contribution to the double OPE representation
(\ref{ope}) of a full conformal block, which consists of
a given primary operator $\O_p$ and all its conformal descendents
$\bigtriangledown^k \O_p$. Let us take the external operators
$\O_i$ in the l.h.s. of (\ref{ope}) to be single--trace, chiral SYM operators
and let us consider the partial waves in which the intermediate
primaries are also single--trace and chiral.
These are the operators
which are in 1--1 correspondence with the single--particle
Kaluza--Klein states of supergravity.
It is then clear that for each
such conformal partial wave one can draw a related
supergravity diagram,
in which the dual KK mode is exchanged in the bulk\footnote{For
a given primary $\O_p$ to contribute
to the double OPE, the 3--point functions
$\< \O_1 \O_3 \O_p \>$  and $\< \O_2 \O_4 \O_p \>$ must be  non--vanishing:
this condition translates on the supergravity side to the existence
of cubic couplings $\phi_1 \phi_3 \phi_p$ and $\phi_2 \phi_4 \phi_p$.},
see Fig.1.

Here we wish to compare supergravity scalar exchange diagrams in
$AdS_{d+1}$
with conformal partial waves in $d$--dimensional
CFT's. We shall find that  for a given partial wave
all the singular  terms in the OPE
are exactly
reproduced by the corresponding
supergravity exchange. However the higher order
terms are different\footnote{This is contrary to the claim in
\cite{liutseytlin2} of an exact
equivalence between supergravity exchanges and conformal partial
waves, but compatible with the results in \cite{liu}. }.

The conformal partial wave for an intermediate
scalar primary is an old result \cite{ferraraopefirst, ferraraope}.
Consider for simplicity the 4--point function
of scalar operator
$\O_i$ with pairwise equal dimensions
($\Delta_1=\Delta_3$, $\Delta_2=\Delta_4$).
Introducing the  conformal invariant variables
\be \label{st}
s  \equiv
\frac{1}{2} \frac{x_{13}^2 x_{24}^2}{x_{12}^2 x_{34}^2 + x_{14}^2
x_{23}^2}\, ,
\qquad \qquad
t  \equiv  \frac{x_{12}^2 x_{34}^2 -x_{14}^2 x_{23}^2}{x_{12}^2 x_{34}^2
+x_{14}^2 x_{23}^2} \,.
\ee
the contribution from an intermediate operator
$\O_\Delta$ and its conformal descendents can be written
as (see Appendix A for the conversion from the form in \cite{ferraraope}
to our notations):

\newpage

\bea \label{partialfinaltext}
\<\O_1(x_1)\O_2(x_2)\O_3(x_3)\O_4(x_4)\>_0 & =&
{1\over x_{13}^{2\D{1}}x_{24}^{2\D{2}}} \,
2{\Gamma\left(\Delta\right)\Gamma\left(\Delta+1-{d\over
2}\right)\over
\Gamma\left({\Delta\over 2}\right)^4}\,
\\
&\times&
\sum_{n=0}^{\infty}s^{\hd+n}\,a_{\hd+n-1}(t)
{\g{\hd+n}^2\over\g{n+\Delta-\d+1}\,n!} \, , \nonumber
\eea
where
\be
a_k(t)=
\sqrt{\pi}{\Gamma(k+1) \over \Gamma\left(k+{3\over 2}\right)}\,
F\left({k+1\over 2},{k\over 2}+1;k+{3\over 2};t^2\right).
\ee
Here all operators are normalized to have unit two--point functions
and the correlators $\<\O_1 \O_3 \O_\Delta\>$
and  $\<\O_2 \O_4 \O_\Delta\>$ are also assumed to have coefficient 1.
Observe that the singular terms in the limit $x_{13} \rightarrow 0$
 are given by $n \leq \Delta_1 -\frac{\Delta}{2} -1$.

The supergravity `t--channel' exchange diagram of Fig.1 in which the
field $\phi$ dual to $\O$ propagates in the bulk can be expressed
in a similar series expansion in terms of the variables $s$ and $t$
(see Appendix A for details). Let $\Delta_1 \leq \Delta_2$.
It is found that in the limit $x_{13} \rightarrow 0$
the terms containing power singularities $O(\frac{1}{x_{13}^k})$ are
\bea
\<\O_{\D{1}}\O_{\D{2}}\O_{\D{3}}\O_{\D{4}}\>\bigg |_{\rm
sing}&=&
{1\over x_{13}^{2\D{1}}x_{24}^{2\D{2}}} \,
2{\Gamma\left(\Delta\right)\Gamma\left(\Delta+1-{d\over
2}\right)\over
\Gamma\left({\Delta\over 2}\right)^4}\,
\\
&\times&
\sum_{n=0}^{\Delta_1 -\frac{\Delta}{2} -1}s^{\hd+n}\,a_{\hd+n-1}(t)
{\g{\hd+n}^2\over\g{n+\Delta-\d+1}\,n!} \, , \nonumber
\eea
in precise agreement with (\ref{partialfinaltext}).
The full series expansion of
the supergravity diagram is however different from
(\ref{partialfinaltext}), for example
logarithmic terms $\sim \log s$ arise at the first non--singular order.
Taking the limit $x_{24} \rightarrow 0$ (keeping the condition $\Delta_1
\leq \Delta_2$) one finds that {\it not}
 all singular powers $O(\frac{1}{x_{24}^k})$
match, but only the terms
more singular than $O(\frac{1}{x_{24}^{2 \Delta_2 -2 \Delta_1}})$.
This is precisely the singularity expected from the contribution to the
OPE of the
double--trace operator $[\O_1 \O_3]$, of approximate dimension
$2 \Delta_1$.
Not surprisingly, the correspondence between conformal partial waves
and AdS exchanges breaks down precisely
when double--traces start to contribute.

We expect similar results for arbitrary spin exchange.
Expressions for conformal partial
waves for arbitrary spin can be found for example in  \cite{ferraraspin}.
The exchange supergravity diagrams
for vectors of general mass and massless gravitons have been evaluated in
\cite{dhfgauge, march99, howto}, where it was also checked that the
leading power
singularity reproduced the contribution expected from OPE considerations
(see (4.23) of \cite{dhfgauge} and Sec. 2.3 of \cite{march99}). It would
be interesting
to extend the comparison to all the subleading power--singular singular terms.

\begin{figure}
\begin{center} \PSbox{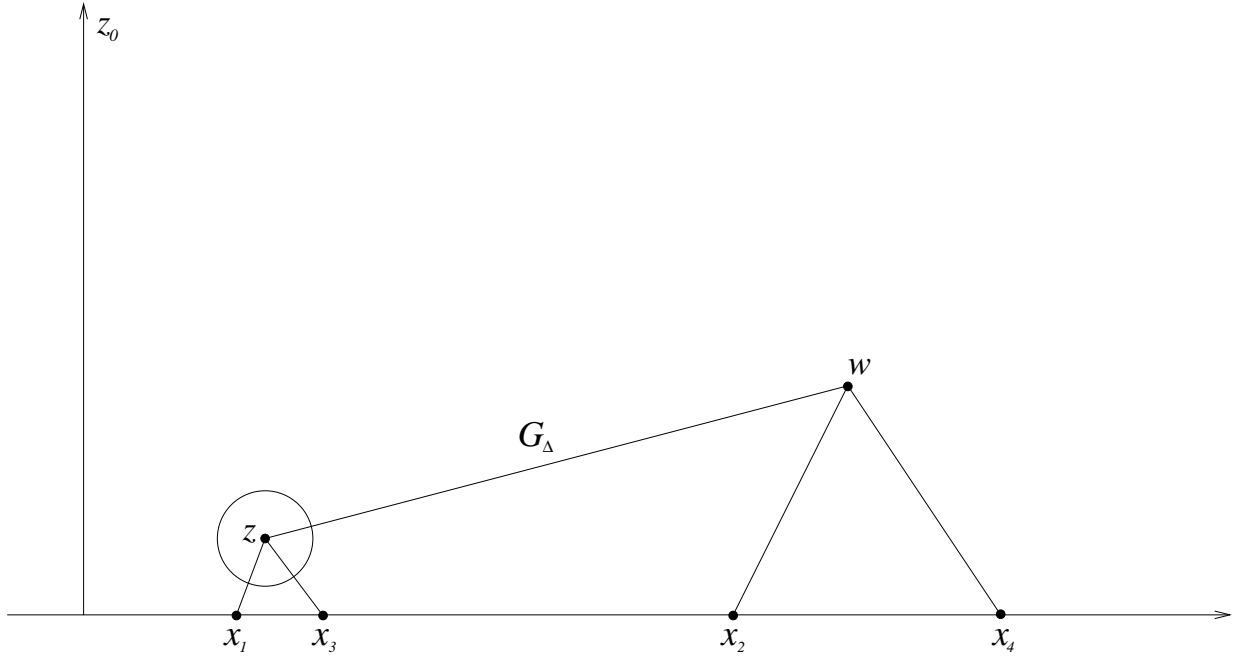 hscale=80 vscale=80}{7.1in}{3.1in} 
\end{center}
\caption{Upper--half plane representation of an exchange integral in
AdS space. In the limit $x_1 \to x_2$ the $z$--integral is
supported in a small ball (in coordinate units) 
close to the boundary point $x_1 \approx x_2$.}
\end{figure}

This nice holographic behavior
of the $AdS_{d+1}$ supergravity exchange diagrams (their  power
singularities match the $d$--dimensional conformal OPE) can also be
understood in the following heuristic way \cite{talk}. The
exchange amplitude is given by an integral over the two bulk interaction
points $z$ and $w$
\be \label{integral} I=
\int [dz] [dw] K_{\Delta_1}(z, \vec x_1)   K_{\Delta_3}(z, \vec x_3)
G_\Delta(z,w)  K_{\Delta_2}(w, \vec x_2)   K_{\Delta_4}(w, \vec x_4)
\ee
where $K$ and $G$ stand for the boundary--to--bulk and bulk--to--bulk
propagators, and $[dz], [dw]$ denote the invariant measures.
Take for concreteness the upper--half plane representation
of AdS
\be
ds^2 = \frac{1}{z_0^2}\left(dz_0^2 + (d\vec z)^2 \right)\,.
\ee
Then
$K_{\Delta}(z, \vec x) \sim \left( \frac{z_0}{z_0^2 +(\vec z-\vec x)^2}
\right)^\Delta$. It is easy to prove that
as we let $\vec x_1 \rightarrow \vec x_3$, the
$z$--integral is dominated by a small coordinate region,
$\left\{ z \;\, {\it s.t.}\, \left[ z_0^2 + (\vec z -\vec x_1)^2\right] \sim x_{13}^2 \right\}$,
which is approaching
the insertion points on the AdS boundary of the two colliding
operators $\O(\vec x_1) \O(\vec x_3)$ (see Fig.2).
This is another example of the UV/IR connection:
short distances in the field theory are probed by large distance physics in the AdS description. As $x_{13} \rightarrow 0$ we can then approximate\footnote{This relation can be
proven by taking  $z_0 \rightarrow 0$ in
the explicit functional form of the normalized $G_\Delta$ as given
for example in (2.5) of \cite{howto}.}
\be \label{replace}
G_\Delta(z,w) \longrightarrow \frac{1}{2 \Delta- d}\;
z_0^\Delta K_\Delta(\vec x_1, w)\, ,
\ee
and we get the expected factorization of (\ref{integral})
into a trivial integral over $z$ and an integral over $w$ which defines the
3--point function $\< \O_\Delta \O_2 \O_4 \>$:
\be
I \longrightarrow \frac{1}{{x_{13}}^{\Delta_1 +\Delta_3 -\Delta}}
\, \< \O_\Delta(\vec x_1) \O_2(\vec x_2) \O_4(\vec x_4) \> \,
\ee
(the numerical coefficients work out exactly). The replacement (\ref{replace}) gives
a clear geometric equivalent, on the supergravity side, of the operator
product expansion $\O_1(\vec x_1) \O_3(\vec x_3) \rightarrow x_{13}^{-\Delta_1 -\Delta_3 +\Delta}
\O_\Delta(\vec x_1)$.
It is possible to compute the first few higher--order
corrections to (\ref{replace}) and to match them exactly \cite{talk}
with the singular contributions
to the operator product $\O_1 \O_3$
of the descendents $\bigtriangledown^k \O_\Delta$.

So far we have analyzed the power singularities in the `direct channel'
limit of an exchange graph, {\it i.e.} when we let approach together two operators
that join to the same bulk interaction vertex ($x_{13} \rightarrow 0$ or
$x_{24} \rightarrow 0$ in (\ref{integral})). A given 4--point function
is obtained by summing all the crossed--symmetric exchanges, as well
as `quartic' graphs (diagrams with a single bulk interaction vertex,
equ.(\ref{D})).
Thus in analyzing the singular behavior
of a 4--point correlator
we also need to consider
the type of leading singularities that appear in quartic graphs, as well as in
the `crossed channel'
limit of an exchange graph (for example $x_{12} \rightarrow 0$ in (\ref{integral})).
It turns out that these two cases (crossed exchanges and quartic)
have the same
qualitative behavior\footnote{These statements can be proved by the methods
reviewed in Appendix A. }: as $x_{12} \rightarrow 0$,
the leading asymptotic
is $O(\frac{1}{x_{12}^{\Delta_1 + \Delta_2 -\Delta_3 -\Delta_4}})$ if
$\Delta_1 + \Delta_2 -\Delta_3 -\Delta_4 > 0$ or  $\log(x_{12})$ if
$\Delta_1 + \Delta_2 -\Delta_3 -\Delta_4 = 0$, and the limit is smooth
if $\Delta_1 + \Delta_2 -\Delta_3 -\Delta_4 < 0$.
These are the singularities
expected from the contribution to the operator product $\O_1 \O_2$
of the composite operator
$[\O_3 \O_4]$, which has
 dimension $\Delta_3 + \Delta_4$ in the large $N$ limit.

The results of this Section have  a clear implication for
the ${\cal N}=4$ SYM theory:
in the limit of large $N$, large $\lambda$, the only singular  terms
in the product of two chiral operators are given by other chiral operators
and their multi--trace products (and their first
few conformal descendents).

\setcounter{equation}{0}
\section{On the Logarithmic Behavior of Conformal Field Theories}

\subsection{General analysis}

Let us analyze the issue of logarithms in general, for an
arbitrary CFT, before returning to the case of the ${\cal N}=4$
supersymmetric Yang--Mills in 4 dimensions.
A CFT is characterized by the absence of any
inherent length
scale. Under quite general conditions one can argue that
for primary operators of
the conformal algebra the 2--point function
is forced by conformal invariance to
have the form
\be
\<{\cal O}(x_1){\cal O}(x_2)\>={C\over |x_1 -x_2|^{2\Delta}}
\label{one}
\ee
The power law on the r.h.s.
 is covariant under scale transformations: if we write $x=\mu x'$,
then the 2--point
 function will be left unchanged provided we also change the operators
  from ${\cal O}$ to ${\cal O}'=\mu ^{-\Delta} {\cal O}$.

One might imagine
 that a {\it logarithmic} dependence $\log{|x_1 -x_2| \over L}$
would violate conformal invariance of the theory, since a length scale $L$ is
needed to make the argument of the logarithm dimensionless. There exists
however
a class of 2d
theories called logarithmic CFTs, where logarithms do arise. These
theories will {\it not}
 be the focus of our interest later on, so we mention them now
and then exclude them from the rest of the discussion below. In logarithmic
CFTs
the dilation operator cannot be diagonalised, but (in the simplest case) has
a Jordan form instead on a
pair of operators ${\cal O}, \tilde{\cal O}$. The 2--point functions are of the form
\bea
\<{\cal O}(z_1){\cal O}(z_2)\>&=&0 \nonumber \\
\<\tilde{\cal O}(z_1){\cal O}(z_2)\>& =& {1\over (z_1-z_2)^{2\Delta}} \\
\<\tilde{\cal O}(z_1)\tilde{\cal O}(z_2)\>&=&
{1\over (z_1-z_2)^{2\Delta}}\log {z_1-z_2\over Y} \nonumber
\label{two}
\eea
(In the above we have considered only the holomorphic
 parts of the operators.)
Under a dilation $z=\mu z'$ we must not only rescale the fields but
also implement a shift transformation:  ${\cal O}' =\mu^{-\Delta} {\cal O},
~\tilde{\cal O}'=
\mu^{-\Delta}[ \tilde{\cal O}+{1\over 2}\log(\mu) {\cal O}]$.
With this change of
variables the correlators of the rescaled theory become identical
to the original ones, and so the
parameter $Y$ in (\ref{two}) does not represent a fundamental length in the
theory, but instead describes a certain choice for the operator
$\tilde{\cal O}$ from
the
subspace of operators
 ${\cal O}, \tilde{\cal O}$ of the same dimension.

These logarithmic CFTs are however not expected to be unitary. In a radial
quantization the dilation operator is the Hamiltonian. In 2--d CFT's the
eigenvalues of the dilation operator on the plane map to the eigenvalues of
the time translation operator
on the cylinder, so in a unitary theory we expect
that
the dilation operator will be diagonalisable and  will not have a
nontrivial
Jordan form. Our case of interest (the 4d ${\cal N}=4$
supersymmetric Yang-Mills theory) is a unitary theory, and we will assume in
what follows that the dilation operator is diagonalisable on the space of
fields. The analysis of the next section will confirm this assumption.

For unitary CFTs the 3--point
function of primary fields also has a standard
form
which is fixed by conformal invariance
\be
\<{\cal O}_1(x_1){\cal O}_2(x_2){\cal O}_3(x_3)\>=
{\gamma_{123} \over x_{12}^{\Delta_1+\Delta_2-\Delta_3}
x_{23}^{\Delta_2+\Delta_3-\Delta_1}x_{13}^{\Delta_1+\Delta_3-\Delta_2}}
\label{three}
\ee
where $x_{ij} \equiv |x_i -x_j|$. 4--point functions are however not fixed in
their functional form by conformal
invariance, though they are restricted
to be a function of two cross ratios 
\bea
&& \<{\cal O}_1(x_1){\cal O}_2(x_2){\cal O}_3(x_3){\cal O}_4(x_4)\>=
{1\over x_{13}^{\S{13}-\D{13}} x_{12}^{\D{13}+\D{24}} x_{14}^{\D{13}-\D{24}}
x_{24}^{\S{24}-\D{13}}}\;
f(\rho,\eta) \\
&&\qquad \qquad \qquad \qquad \eta={x_{12}^2x_{34}^2\over x_{13}^2x_{24}^2}, \quad
\rho={x_{14}^2x_{23}^2\over
x_{13}^2x_{24}^2}
\label{six}
\eea
where $\Sigma_{ij} = \Delta_i +\Delta_j$, $\Delta_{ij}= \Delta_i-\Delta_j$.
One might imagine that the expansion of $f$ in say $\eta$ would in general
contain a logarithm:
\be
f(\eta,\rho)=\dots+f_{-1}(\rho)\eta^{-1}+f_0(\rho)+\tilde f_0(\rho)\log \eta +
f_1(\rho)\eta+\dots
\label{seven}
\ee
so that the $x_1\rightarrow x_2$ limit of the 4-point function would contain
a logarithmic term $\sim \log x_{12}$. But consider evaluating the 4--point
function using the OPE:
\bea
{\cal O}_1(x_1){\cal O}_2(x_2)&=&\sum_p {\cal O}_p(x_2) x_{12}^{\Delta_p-\Delta_1-\Delta_2}
\nonumber \\
{\cal O}_3(x_3){\cal O}_4(x_4)&=&\sum_q {\cal O}_q(x_4) x_{34}^{\Delta_q-\Delta_3-\Delta_4} \label{3ope}
\\
\<{\cal O}_1(x_1){\cal O}_2(x_2){\cal O}_3(x_3){\cal O}_4(x_4)\>&=&\sum_{p,q} \<{\cal O}_p(x_2){\cal O}_q(x_4)\>
x_{12}^{\Delta_p-\Delta_1-\Delta_2}
x_{34}^{\Delta_q-\Delta_3-\Delta_4} \nonumber
\label{eight}
\eea
We have written the functions arising in the OPE in symbolic form: the
operators
${\cal O}_p$ will
in general carry tensor indices and these can contract with the unit
vector
along $x_1-x_2$.
But the basic point that we wish to observe is the following.
If ${\cal O}_p$ is a conformal primary, then the coefficient function
appearing in
the
OPE of ${\cal O}_1$ and ${\cal O}_2$ can be deduced from the 3--point
function
(\ref{three}), and contains no logarithm.
If ${\cal O}_p$ is a conformal descendant,
then the coefficient function will be obtained as derivatives of the
coefficient
function for the corresponding primary, and so again there will be no logarithm
in these functions. Similar arguments yield that if the 2-point functions of
primaries contain no logarithms, then nor do the two point functions
$\<{\cal O}_p(x_2){\cal O}_q(x_4)\>$ appearing in the last line of the above equation.

Thus if the OPE sums in (\ref{3ope}) converge, we  conclude
that the 4--point function does not in fact have a logarithm in $x_{12}$ in
an expansion around $x_{12}=0$. In unitary 2--d CFTs such OPE sums do yield
the correct 4--point functions, and logarithms do not arise in the short
distance expansions.

Let us now consider the circumstances where we {\it will} encounter logarithms
in our analysis of CFT correlation functions.
Suppose  that we study a 1--parameter family of CFTs;
let us denote this parameter by $a$.
 Suppose that the theory at a particular value of the parameter, say zero,
  is particularly simple.
  Then we may ask for the $n$--point correlation functions in a series in the
  parameter around zero. The example that we
   are concerned with is of course ${\cal N}=4$ $SU(N)$ SYM,
   where the theory for $1/N\rightarrow 0$ is expected to be
simple\footnote{We are actually 
interested in the double limit $N \to \infty$,
$\lambda \to \infty$. Since in the supergravity analysis there is no
remnant of the $\lambda$ dependence, while the $1/N$ expansion
coincides with the perturbative expansion in powers of the Newton constant,
$G_5 \sim 1/N^2$, we restrict here to the $N$ dependence alone.},
and the correlators may be studied in a series in $1/N$.
Consider first the 2--point function (\ref{one}). Let
\be
\label{hthree}{\Delta=\Delta^{(0)}+\Delta^{(1)} }
\ee
where $\Delta^{(1)}$ vanishes as  $a\rightarrow 0$. Then we may write
\bea
\label{onep}
\<{\cal O}(x_1){\cal O}(x_2)\>&=&{C\over x_{12}^{2\Delta^{(0)}
}}\frac{1}{L^{2\Delta^{(1)}}}
\left( {L \over x_{12}} \right)^{2\Delta^{(1)} }\\
&& \approx {C \over x_{12}^{2\Delta^{(0)} }}
\frac{1}{L^{2\Delta^{(1)}} }
\left( 1-2\Delta^{(1)}\log{ x_{12} \over L } \right)
\eea
Here we have introduced a length scale $L$ to be able to write the
logarithms in dimensionless form: $L$ may be chosen in an arbitrary way
since it just defines the normalisation of the operators,
and is needed because the operators ${\cal O}$ have at a generic value of $a$
a dimension
that is different from the one at $a=0$.

Thus we see that the leading  correction to the 2--point function
at nonzero $a$  has a
logarithm in $x_{12}$, with a coefficient that depends
in the manner shown on the correction $\Delta^{(1)}$
 to the dimension at nonzero
$a$.
 Such a correction will of course
 be absent if the dimensions of the operators do not change with $a$. Thus
 in particular for our case of interest, the two point functions of
SYM chiral operators will not have such a logarithmic correction.
But we can consider
 composite operators made from two chiral operators, and such operators will
 in general have a logarithm in their 2--point function. We will compute the
 4--point function of four chiral operators; if we then take the limit
 $x_{12} \rightarrow 0$, $x_{34}
 \rightarrow 0$, then we can extract the
 2--point function of the nonchiral composites, and observe the logarithmic
 correction.

Now consider the 3--point function.
Suppose for simplicity that the operators
${\cal O}_1$ and ${\cal O}_2$ have dimensions that are protected (i.e. unchanged under
variations of the parameter $a$). Let the dimension of ${\cal O}_3$ be
$\Delta_3=\Delta^{(0)}+\Delta^{(1)} $. Then we have
\bea
\label{hfour}
\<{\cal O}_1(x_1){\cal O}_2(x_2){\cal O}_3(x_3)\>&=&{\gamma_{123}\over
x_{12}^{\Delta_1+\Delta_2-\Delta^{(0)}}x_{23}^{\Delta_2+\Delta^{(0)}-\Delta_1}
x_{13}^{\Delta^{(0)}+\Delta_1-\Delta_2}} \nonumber
\\
&& \cdot \frac{1}{L^{2\Delta^{(1)}}}\left(
1-\Delta^{(1)}\log{ x_{13}x_{23}\over x_{12}\,L
} \right)
\eea
 As mentioned above, what we will do is compute a 4--point function of chiral
 oprators, and the  composite operator appearing in the above equation
 will be obtained in a limit where two of the chiral insertions are taken to
 approach each other.

In the above we have considered the case of operators that change
by  a multiplicative factor under scale transformations.
In a unitary CFT we
can always find a basis of operators where the action of the dilation
operator is thus diagonalised. But if we are considering a one parameter
family of theories, then operators which are
dilation eigenstates
at say $a=0$ will not generically be such in the theory with $a \neq 0$.
We will be 
interested in computing only the leading
order correction to the dimensions, so 
 we can neglect mixing among operators
that have different dimensions at $a=0$ (this is a familiar
story, for example from ordinary quantum mechanics perturbation
theory), and we need to consider only mixing among operators
which are degenerate at zeroth order. 
Let us consider the logarithms arising
at the first order in correction away from $a=0$, in the case where
there are two or more degenerate operators at $a=0$.

At this point for clarity of the discussion we specialize to the
case of the theory that we are going to study -- the ${\cal N}=4$
supersymmetric
$SU(N)$ Yang-Mills theory, studied in a $1/N$ expansion around the
large $N$ limit.

Let ${\cal O}_i$  be single--trace
chiral operators ({\it i.e.}, chiral primaries
${\rm tr} X^{(i_1} \cdots X^{i_k)}$ or any of their supersymmetry descendents).
  Superconformal symmetry
 fixes their scaling dimension to a value independent of $N$.
    Recall that these operators are a small subset of all the operators that
are
  primaries of the conformal algebra.
   We take the ${\cal O}_i$ to be normalized
such that
\be
{\cal O}_i(x){\cal O}_j(0)={\delta_{ij}\over |x|^{2\Delta_i}}
\ee
We now construct the double--trace  composite operators made
from pairs of these primaries.
Let ${\cal O}_{ij}$ be  given through
\be
[
{\cal O}_{ii}(x)
]_y \equiv  {1\over \sqrt{2}} {\cal O}_i(x+y)
{\cal O}_i(x)
-{\rm subtractions}\ee
\be
[
{\cal O}_{ij}(x)
]_y  \equiv  {\cal O}_i(x+y){\cal O}_j(x) -{\rm subtractions},
 ~~(i\ne j)
\ee
To define these composite operators we point--split the two chiral
constituents by a distance
$y$ and subtract the power--singular contributions
in their OPE. In the supergravity limit of $N \to \infty$ 
and $\lambda$ large
the singular part of the OPE of two single--trace
chiral operators ${\cal O}_1$ and ${\cal O}_2$ is given
by other single--trace chiral operators with dimension 
$\Delta <\Delta_1+\Delta_2$
(with an OPE  coefficient $O(\frac{1}{N})$)
and
by double--trace composites $[{\cal O}_3 {\cal O}_4]$
with $\Delta_3 + \Delta_4 < \Delta_1 + \Delta_2$ 
(with an OPE coefficient
$O(\frac{1}{N^2}))$. 
These statements follow from standard large $N$
counting rules and from the input of the AdS/CFT duality (confirmed
by the analysis of Section 2) that non--chiral operators dual to string
states have a large anomalous dimension in this limit.

  Let us denote the composite indices
$\{ij\}$ by $\alpha, \beta\dots$. Also, to simplify the notation
we will usually drop the explicit dependence on the cut--off distance $y$
and simply indicate the double--trace composites
by ${\cal O}_\alpha$.
Since the dimensions of the
chiral operators are integral, there are clearly several composites for which
the sum of the dimensions of the chiral constituents equals the same value.
For $N \to \infty$ all these operators will have the same dimension.
Let us consider the subspace of composites $\{ {\cal O}_\alpha \}$
that have the same dimension $\Delta$ in the large $N$ limit,
 $\Delta_\alpha =
\Delta_i +\Delta_j= \Delta$ for all $\alpha$.
At $1/N=0$ the 2--point function of the composites is obviously
\be
\<{\cal O}_\alpha(x){\cal O}_\beta(0)\> =
{\delta_{\alpha\beta}\over |x|^{2\Delta}}
\ee
In the above we have assumed that $y \ll x$.

For $1/N$ small but non--zero, these composites will mix
among each other under scale tranformations. The infinitesimal
dilation operator is simply $y \frac{d}{dy}$, so we have
\be \label{ymixing}
y \frac{d}{dy} {\cal O}_\alpha = M_{\alpha \beta} {\cal O}_\beta
\ee
It will be apparent from the supergravity analysis of the next
section that $ M_{\alpha \beta}$ is a real symmetric matrix, as expected
in a unitary theory. We can then find the dilation eigenstates and
their anomalous dimensions by solving the eigenvalue problem for $M$. Let
the eigenvectors of the dilation $y \frac{d}{dy}$
be the operators ${\cal O}_{(A)}$
\be
{\cal O}_{(A)}=V_{(A)}^\alpha {\cal O}_\alpha; ~~{\cal O}_\alpha=V^{(A)}_\alpha {\cal O}_{(A)}
\ee
\be
y \frac{d}{dy} {\cal O}_{(A)} = \Delta_{(A)} {\cal O}_{(A)}\,.
\ee
Composite operators $ [{\cal O}_{(A)}]_y$ defined 
with a certain value of the cut--off are related to
the composites with a different cut--off $y'$ by a simple
rescaling
\be
 [{\cal O}_{(A)}]_y= \left(  \frac{y}{y'} \right)^{\Delta_{(A)}}
 [{\cal O}_{(A)}]_{y'} \,.
\ee
The orthogonal matrix
\be
(\hat V)_\alpha{}^{(A)}=V^{(A)}_\alpha
\ee
accomplishes the change of basis that makes $M$ diagonal
\bea \label{MD}
M_{\alpha\beta}& =& (\hat V \hat D  \hat V {}^T)_{\alpha\beta} \\
\hat D& =& {\rm diag}\{\Delta_{(1)}, \dots \Delta _{({n})}\}
\eea
The full conformal dimension of ${\cal O}_{(A)}$ is 
\be
\Delta + \Delta_{(A)}, ~~\Delta_{(A)} \ll \Delta
\ee
From the supergravity results (or from field--theory large $N$ counting)
we have that $\Delta_{(A)} = O(\frac{1}{N^2})$.
Since conformal invariance implies that primary operators with different
dimensions are orthogonal, we have
\be
\<{\cal O}_{(A)}(x){\cal O}_{(B)}(0)\>_a=|y|^{2 \Delta_{(A)}}{\delta_{AB}\over
|x|^{2\Delta+2 \Delta_{(A)}}}
\ee
where the power of $y$ reflects the fact that we have chosen to normalize
the  composite operators at scale $y$, for all values of the parameter
$1/N$. We then get
\bea \label{long}
&& \<{\cal O}_\alpha(x){\cal O}_\beta(0)\>_a=V_\alpha^{(A)}V_\beta^{(B)}\<{\cal O}_{(A)}(x){\cal O}_{(B)}(0)\>_a  \\
&& =V_\alpha^{(A)}V_\beta^{(B)}|y|^{2\Delta_{(A)}}
\frac{\delta_{AB}}{|x|^{2\Delta+2\Delta_{(A)}}}
\approx V_\alpha^{(A)}V_\beta^{(B)}\frac{\delta_{AB}}{|x|^{2\Delta} }
\left[1- 2 \Delta_{(A)}\log \left(\frac{x}{y}\right)\right] \nonumber
\eea
We see that if we have the 4--point function of chiral operators
(assume $\Delta_i + \Delta_j = \Delta_k + \Delta_l = \Delta$)
\be \label{chirallog}
\<{\cal O}_i(x+y){\cal O}_j(x){\cal O}_k(y'){\cal O}_l(0)\>, ~~y,y'\ll x
\ee
then we get to order $O(\frac{1}{N^2})$  a term
$$\frac{1}{x^{2\Delta}}\,\log \frac{y y'}{x^2}$$ with coefficient
\be
\sum_{(A)}V_{\alpha}^{(A)}V_{\beta}^{(A)}\Delta_{(A)}= M_{\alpha \beta}, ~~\alpha=(ij),
\beta=(kl)
\ee
Thus from the knowledge of the leading logarithmic
term in the 4--point functions of chiral operators we can directly
extract the mixing matrix $M_{\alpha \beta}$ of the double--trace
composites, and by solving
the eigenvalue problem we can then find the dilation eigenstates and their
anomalous dimensions. Observe that for $\alpha = \beta$, {\it i.e.}
when the chiral constituents are pairwise equal (say  $i=k$, $j=l$) the
logarithmic term in (\ref{long}) appears as a $O(\frac{1}{N^2})$
correction to the leading power behavior, which is of order $O(1)$
in the large $N$
counting and is given in (\ref{chirallog}) by the disconnected
contribution to the 4--point function (obtained by simply
contracting the equal chiral operators). However
for $\alpha \neq \beta$, {\it i.e.} in the case
of mixing between different composites, the log
naturally appears
to order $O(\frac{1}{N^2})$ without a $O(1)$ power term.

Let us finally relate this discussion to the Operator Product Expansion.
From the very definition of the composite double--trace operators, we
have
\be
{\cal O}_i(x) {\cal O}_j(0)  \sim \;{\rm singular\; terms} \; +
[{\cal O}_\alpha(0)]_x \; , ~~\alpha=(ij)
\ee
We can rewrite
\bea
&&[{\cal O}_\alpha]_x = V_{\alpha}^{(A)}\,[{\cal O}_{(A)}]_x
=  V_{\alpha}^{(A)}\,
\left({x}/{y}\right)^{\Delta_{(A)}}
[{\cal O}_{(A)}]_y =  V_{\alpha}^{(A)}\,
\left({x}/{y}\right)^{\Delta_{(A)}}
V_{(A)}^\beta\, [{\cal O}_{\beta}]_y\\
&&\approx \left[   V_{\alpha}^{(A)}  V_{(A)}^\beta +  V_{\alpha}^{(A)}
\Delta_{(A)}   V_{(A)}^\beta \, \log (x/y)              \right]
 [{\cal O}_{\beta}]_y =[{\cal O}_\alpha]_y + M_{\alpha \beta} \,
\log (x/y) [{\cal O}_\beta]_y \nonumber
\eea
where in the last step we have used  the orthogonality
of the matrix  $\hat V$  and the relation (\ref{MD}). Thus to order
$O(\frac{1}{N^2})$ the OPE takes the form
\be \label{opemixing}
{\cal O}_i(x) {\cal O}_j(0)  \sim \;{\rm singular\; terms} \;
+ [{\cal O}_{\alpha}(0)]_y + M_{\alpha \beta}
\log ({x}/{y})\; [{\cal O}_\beta(0))]_y \,,~~~\alpha=(ij)
\ee
This equation translates in the OPE language the renormalization
and mixing of the double--trace composites. This is the form
of the OPE required by compatibility
with the action  of the dilation operator. In fact,
the l.h.s. of the above OPE is clearly independent
from the scale $y$, and
applying $y \frac{d}{dy}$ to the r.h.s. it is immediate
 to see that we get
identically  zero
 (to order $O(\frac{1}{N^2})$)
once the mixing  (\ref{ymixing})
is taken into account.

\subsection{Logarithms in the supergravity picture}

\begin{figure}
\begin{center} \PSbox{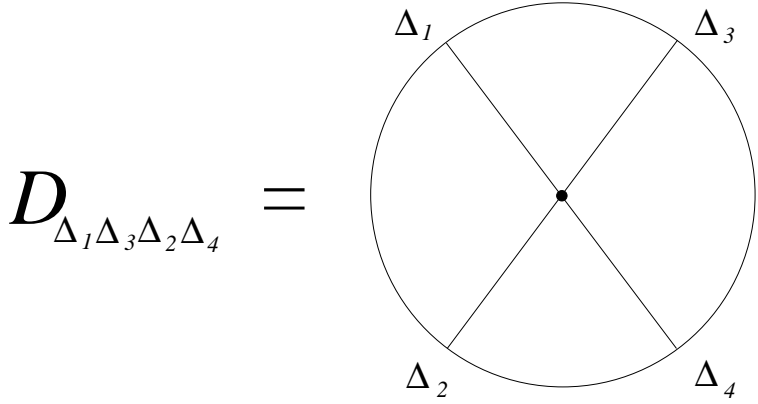 hscale=80 vscale=80}{5in}{2.0in} 
\end{center}
\caption{Quartic graph.}
\end{figure}

The AdS/CFT correspondence provides an interesting dual picture of the
Yang-Mills theory, and in this dual picture we have a simple and elegant
pictorial way of seeing the appearance of the above logarithms. 
Consider
a 4--point function $\<{\cal O}(x_1){\cal O}(x_2){\cal O}(x_3)
{\cal O}(x_4) \>$.
At $1/N=0$,
the supergravity theory is free, and the 4--point function
factorizes into the product of the 2--point functions obtained
by pairwise contractions of the operators.
At order $O(\frac{1}{N^2})$,  we get contributions
from tree--level supergravity graphs, either
of the exchange type (with two cubic vertices, as  for example Fig.1) 
or quartic graphs with a single interaction vertex, as in Fig.3.
 These connected graphs include logarithmic corrections to the correlator,
which represent
corrections to the composite operators generated by the 
approach of two chiral
primaries.
 To see that the logarithm indeed arises from the vicinity of the
composite operator in the supergravity diagram, we look at a typical 
term that
arises in the supergravity description. Consider for simplicity the quartic
graph, Fig.3. 
The AdS space is represented as the upper half space in Fig.4.
The quartic vertex is at a location $z$, which must be integrated over. In
Fig.4 we have partitioned the
domain of integration of $z$ into annular regions, each of which is
$e$ times  in
diameter compared to the one nested inside it. The integration has the form
(take for simplicity all conformal dimensions to be equal)
\be
\int {d^5z\over z_0^5}{z_0^\Delta\over (z-x_1)^{2\Delta}}
 {z_0^\Delta\over (z-x_2)^{2\Delta}}
 {z_0^\Delta\over (z-x_3)^{2\Delta}}
 {z_0^\Delta\over (z-x_4)^{2\Delta}}
\ee

Let $x_1, x_2$ be close to the point $x=0$,  and consider the region of
integration where $z$ approaches $z=0$ as well.  The
points $x_3, x_4$ are assumed to be far away from this region. Then we may
approximate
\be
{z_0^\Delta\over (z-x_3)^{2\Delta}}
 {z_0^\Delta\over (z-x_4)^{2\Delta}}\approx { z_0^{2\Delta}\over
 x_3^{2\Delta}x_4^{2\Delta}}
\ee
 If we set $x_1, x_2$ to zero the integral becomes
\be
{1\over
 x_3^{2\Delta}x_4^{2\Delta}}\int {d^5z\over
z_0^5}z_0^{-2\Delta}z_0^{2\Delta}
\ee
 which has an equal contribution from each annulus in Fig.4, and so can be
 seen
 to diverge logarithmically. The actual  integral we have  is cut off at
$x_1-x_2$ in the
 UV, and at $\sim x_1-x_3$ in the IR, and so is of order
$$
\sim \log {x_{13}\over x_{12}}$$
 which is a logarithm of the kind that we will
 observe in the supergravity diagrams. In more general integrals we will have
terms with singular powers in $x_{12}$ as 
well\footnote{When a single variable $z$ is being integrated and the leading 
singularity is more singular than a logarithm then this singularity 
is of the form $\sim {1\over x_{12}^k}$; it cannot be of the form 
$\sim {1\over x_{12}^k}\log x_{12}$. Logarithms can however 
appear at subleading orders in $x_{12}$.}. Further the 
exchange graphs like that
in
Fig.1 have two vertices $z$, $w$ that are integrated over, and
a contribution $\log |x_{12}|$  can arise when one or both of these vertices
are in the vicinity of $x_1, x_2$.
\begin{figure}
\begin{center} \PSbox{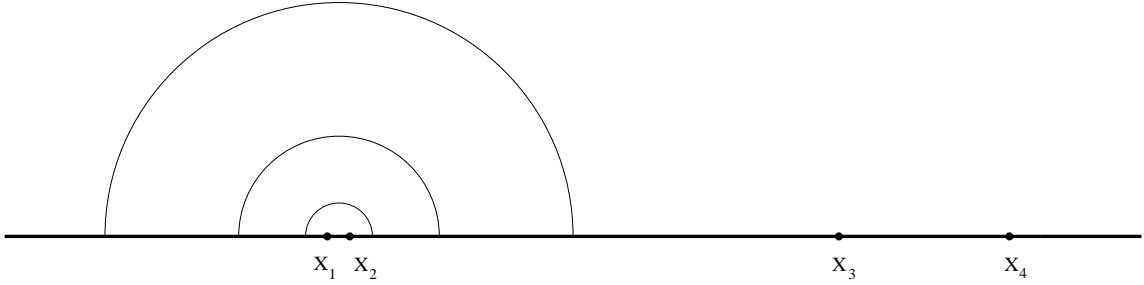 hscale=70 vscale=70}{6.1in}{1.4in} 
\end{center}
\caption{A quartic graph in the upper--half plane representation of
AdS space. Each annulus is $e$ times bigger in diameter 
than the one nested inside it. The $z$--integration receives an
equal contribution from each annular region.}
\end{figure}

\setcounter{equation}{0}
\section{Dilaton--Axion Four Point Functions and Anomalous
Dimensions}

In this section we will consider the supergravity results \cite{march99}
for 4--point functions of the SYM
operators $\ophi$, $\oc$ dual to the dilaton and axion fields.
Following the logic of the previous section, we can extract
information about the $O(\frac{1}{N^2})$
 anomalous dimensions and mixings of the
operators occurring in the OPE of $\ophi$ and $\oc$
by looking at the logarithmic behavior of the correlators.

The expressions for the 4--point functions
of normalized operators (see (\ref{norm})), for $\lambda \to \infty$
and up to order $O(\frac{1}{N^2})$,  are summarized in
Appendix C, equations (\ref{4dil}) and (\ref{fulldilax}).
The disconnected graphs
give some trivial  powers of the separation, of order $O(1)$ in the
large $N$ counting,
 while the connected tree--level
supergravity graphs
provide $O(\frac{1}{N^2})$ contributions which are non--trivial
functions of the cross--ratios.
The leading logarithmic asymptotics are
given in (\ref{4dillog}--\ref{dilaxs}). They have the structure expected
from the contribution to the OPE of double--trace composites of
dimensions $8+O(\frac{1}{N^2})$.

Let us first discuss the simplest
case, the  `s--channel' limit ({\it i.e.}, $x_{12} \to 0$, $x_{34} \to 0$)
of the correlator $\< \hat \ophi(x_1)  \hat \oc(x_2)  \hat \ophi(x_3)  \hat \oc(x_4) \>$.
In this limit there are no power singularities and the logarithmic
term in (\ref{dilaxs}) is the leading contribution.
Thus from (\ref{fulldilax}--\ref{I0p}) and
(\ref{dilaxs}) we obtain one of the leading coefficient
functions of the OPE of $\hat \ophi$ and $\hat \oc$
\be \label{phic}
\hat \ophi (x_1) \hat \oc (x_2) = A_{\phi c} (x_{12}, y)
[\hat \ophi \hat \oc]_y(x_2) + \cdots
\ee
Here, $[\hat \ophi \hat \oc]_y$ is the composite (double--trace) operator
defined by the above equation with
$A_{\phi c}$ given to order $O(\frac{1}{N^2})$ and for
$\lambda \to \infty$ by
\be
A_{\phi c} (x_{12}, y) = 1 + {A \over N^2} \ln \left(
\frac{x_{12}}{y} \right) .
\ee
The  numerical constant $A$ is readily determined
from the logarithmic asymptotics as $x_{12} \to 0$ of
 $\< \hat \ophi  \hat \oc  \hat \ophi  \hat \oc \>$ given by (\ref{dilaxs}):
\be
A= -16\,.
\ee
The above leading behavior of the coefficient function
receives  corrections both in inverse powers of $\lambda$ and
in the $\frac{1}{N}$ expansion. For example, tree--level
stringy corrections, of order $O(\alpha'^3)$ with respect
to the Einstein--Hilbert action, give a
$O(\frac{1}{N^2 \lambda^{3/2}})$ contribution. The first quantum
corrections (one loop in supergravity) are of order $O(\frac{1}{N^4})$.

Next, from the `t--channel' limit ($x_{13} \to 0$, $x_{24} \to 0$)
of the same correlation function
$\langle \hat \ophi  \hat \oc \hat \ophi \hat \oc \rangle$, as well as of the
correlators $\langle \hat \ophi \hat \ophi \hat \ophi \hat \ophi \rangle$ and
$\langle \hat \oc \hat \oc \hat \oc \hat \oc \rangle$,
we can extract terms in the OPE of two $\hat \ophi$'s and two
$\hat \oc$'s.
We expect on general grounds that the OPE will assume the schematic
form
\bea \label{ope2phi}
\hat \ophi (x_1) \hat \ophi (x_3) & = & {I \over x_{13}^8} +
{T \over x_{13}^4}
+ {\partial  T \over x_{13}^3} + \dots
+ C_{\phi \phi} \ [\hat \ophi \hat \ophi]_y
+ C_{\phi c}  [\hat \oc \hat \oc]_y
+ C_{\phi T}  [T T]  + \cdots \qquad
\\
\hat \oc (x_2) \hat \oc (x_4) & = & {I \over x_{24}^8} + {T \over x_{24}^4}
+ {\partial  T \over x_{24}^3} + \dots
+ C_{c \phi} \ [\hat \ophi \hat \ophi]_y
+ C_{c c}  [\hat \oc \hat \oc]_y
+ C_{c T}  [T T]  + \cdots \qquad
\eea
We have suppressed for the sake of brevity
the dependence of the coefficient functions
upon the positions $x_i$, and the Lorentz structures in the
stress--energy tensor terms. Unlike the $\hat \ophi \hat \oc$ OPE,
here there are  some power--singular terms, arising  from the contributions
of the stress--energy tensor and its first descendents. We discussed
these terms in Section 2, where we checked
that the singular powers of the `direct channel' supergravity exchange graph
(in this case, the t--channel graviton exchange)
exactly match the predictions of the OPE\footnote{Here we have a tensor
rather than a scalar exchange, but we expect a completely analogous result.}.

Let us then analyze the contributions to the above OPE's of the double--trace
composites. Clearly, the
correlator $\langle \hat \ophi \hat \oc \hat \ophi \hat \oc \rangle$
determines the coefficient functions $C_{\phi c}=C_{c \phi}$,
while $\langle \hat \ophi \hat \ophi \hat \ophi \hat \ophi \rangle$
determines
$C_{\phi \phi}$ and $\langle \hat \oc \hat \oc \hat \oc \hat \oc \rangle$
determines $C_{cc}$.
Since $\langle \hat \ophi \hat \ophi \hat \ophi \hat \ophi \rangle =
\langle \hat \oc \hat \oc \hat \oc \hat \oc \rangle$,
we immediately have that
$C_{\phi \phi}=C_{cc}$.
Thus,
\bea \label{Cphiphi}
C_{\phi \phi} = C_{cc} & = &
  1 + {C \over N^2} \ln \left( \frac{x_{13}}{y} \right)
\\
\label{Cphic}
C_{\phi c} = C_{c \phi } & = &
  1 + {B \over N^2} \ln \left( \frac{x_{13}}{y} \right) \,.
\eea
From the logarithmic asymptotics (\ref{4dillog}, \ref{dilaxt})
of the 4--point functions,
 we find that\footnote{The constant $C$ is a half of the coefficient
in (\ref{4dillog}) because there are 2 possible Wick contractions
of the composite $[\hat \ophi \hat \ophi](x_3)$ generated in the
OPE (\ref{ope2phi}) with
the two remaining operators $\ophi(x_2)$ and $\ophi(x_4)$.}
\be
B= \frac{2^7}{3\cdot 7}
\qquad \qquad
C = -\frac{2^4 \cdot 13}{3 \cdot 7}\,.
\ee

Now, it is clear from the above OPE's and identification of the
coefficient functions that we lack some information on the
data for operators in the OPE whose dimension is approximately 8.
For example,  to compute $C_{cT}$ and $C_{\phi T}$ would require
knowledge of correlators
$\langle T_{\mu \nu}  T^{\mu \nu} \ophi \ophi \rangle $
and $\langle T_{\mu \nu}T^{\mu \nu}  \oc \oc \rangle$,
which are not at present available. The correlator $\langle TTTT \rangle$,
which is even more out of reach, would also be necessary.
 In order to extract
the dilation eigenstates and their anomalous
dimensions, we would need to compute the full mixing
matrix of the operators of dimension approximately 8 in
the large $N$, large $\lambda $ limit. There are actually
several such operators that we omitted when writing down the OPE's
above. The reason is that
many more operators of dimension  8 in the free--field approximation
may be constructed by taking the product
of conformal descendants, such as $[\partial J \partial J]
= [\O^{(3)}_0 \O^{(3)}_0]$, or even products of fermion operators.
Clearly, obtaining the
full mixing matrix is possible but very involved,
since it would require computing several 4--point functions in supergravity.

Fortunately, using the invariance of the theory under the
$U_Y(1)$ symmetry in the supergravity limit,
it is possible to disentangle the OPE and isolate some operators that
mix in a simple way. This is done in the next section.
\begin{table}
\begin{center}
\begin{tabular}{|l|c|c|c|c|c|c|c|} \hline 
SYM Operator   & desc & SUGRA & dim & spin  & $Y$ & $SU(4)$ &
 lowest reps
                \\ \hline \hline
$\O _k         \sim   \tr X^k$, $k\geq 2$        
                  & -- 
                  & $h_{\alpha}^{\alpha}\ \ a_{\alpha \beta \gamma \delta}$ 
                  & $k$
                  & $(0,0)$    
                  & 0 
                  & $(0,k,0)$
                  & {\bf 20'},{\bf 50},{\bf 105}
                \\ \hline
$\O _k ^{(1)}    \sim \tr \lambda X^k$, $k\geq 1$
                  & $Q$  
                  & $\psi _{(\alpha)}$ 
                  & $k+\32$
                  & $(\12, 0)$             
                  & $\12$ 
                  & $(1,k,0)$
                  & {\bf 20},{\bf 60},{\bf 140'}
                \\ \hline
$\O _k ^{(2)}    \sim \tr \lambda \lambda X^k$ 
                  & $Q^2$  
                  & $A_{\alpha \beta} $
                  & $k+3$
                  & $(0,0)$ 
                  & $1$ 
                  & $(2,k,0)$
                  & {\bf 10$_c$},{\bf 45$_c$},{\bf 126$_c$}
                \\ \hline
$\O _k ^{(3)}   \sim  \tr \lambda \bar \lambda X^k$   
                  & $Q \bar Q$
                  & $ h_{\mu \alpha} \ \ a_{\mu \alpha \beta \gamma}$
                  & $k+3$
                  & $(\12,\12)$ 
                  & $0$ 
                  & $(1,k,1)$
                  & {\bf 15},{\bf 64},{\bf 175}
                \\ \hline
$\O _k ^{(4)} \!  \sim \! \tr F_+ X^k$, $k\geq 1$    
                  & $Q^2$ 
                  & $A_{\mu \nu}$
                  & $k+2$
                  & $(1,0)$ 
                  & $1$ 
                  & $(0,k,0)$
                  & {\bf 6$_c$},{\bf 20$_c$},{\bf 50$_c$}
                \\ \hline
$\O _k ^{(5)}   \sim  \tr F_+ \bar \lambda X^k$ 
                  & $Q^2\bar Q$
                  & $\psi _\mu$
                  & $k+\72$
                  & $(1, \12)$ 
                  & $\12$ 
                  & $(0,k,1)$
                  & ${\bf 4}^*,{\bf 20}^*,{\bf 60}^*$
                \\ \hline
$\O _k ^{(6)}    \sim \tr F_+ \lambda X^k$ 
                  & $Q^3$
                  & ``$\lambda$"
                  & $k+\72$
                  & $(\12 ,0)$ 
                  & $\32 $ 
                  & $(1, k,0)$
                  & {\bf 4},{\bf 20},{\bf 60}
                \\ \hline
$\O _k ^{(7)}   \sim  \tr \lambda \lambda \bar \lambda X^k$ 
                  & $Q^2\bar Q$
                  & $\psi _{(\alpha)}$
                  & $k+\92$
                  & $(0,\12)$ 
                  & $\12$ 
                  & $(2,k,1)$
                  & {\bf 36},{\bf 140},{\bf 360}
                \\ \hline
$\O _k ^{(8)}    \sim \tr F_+^2 X^k$   
                  & $Q^4$
                  & $B$
                  & $k+4$
                  & $(0,0)$ 
                  & $2$ 
                  & $(0,k,0)$
                  & {\bf 1$_c$},{\bf 6$_c$},{\bf 20'$_c$}
                \\ \hline
$\O _k ^{(9)}   \sim \tr F_+ F_- X^k$ 
                  & $Q^2\bar Q^2$
                  & $h_{\mu \nu}'$
                  & $k+4$
                  & $(1,1)$ 
                  & $0$ 
                  & $(0,k,0)$
                  & {\bf 1},{\bf 6},{\bf 20'}
                \\ \hline
$\O _k ^{(10)}   \sim \tr F_+ \lambda \bar \lambda X^k$ 
                  & $Q^3\bar Q$
                  & $A_{\mu \alpha}$
                  & $k+5$
                  & $(\12,\12)$ 
                  & $1$ 
                  & $(1,k,1)$
                  & {\bf 15},{\bf 64},{\bf 175}
                \\ \hline
$\O _k ^{(11)}   \sim \tr F_+ \bar \lambda \bar \lambda X^k$ 
                  & $Q^2\bar Q ^2$
                  & $a_{\mu \nu \alpha \beta}$
                  & $k+5$
                  & $(1, 0)$ 
                  & $0$ 
                  & $(0,k,2)$
                  & {\bf 10$_c$},{\bf 45$_c$},{\bf 126$_c$}
                \\ \hline
$\O _k ^{(12)}   \sim \tr \lambda \lambda \bar \lambda \bar \lambda X^k$ 
                  & $Q^2\bar Q ^2$
                  & $h_{(\alpha \beta)}$
                  & $k+6$
                  & $(0, 0)$ 
                  & $0$ 
                  & $(2,k,2)$
                  & {\bf 84},{\bf 300},{\bf 2187}
                \\ \hline
$\O _k ^{(13)}   \sim \tr F_+^2 \bar \lambda X^k$ 
                  & $Q^4\bar Q$
                  & ``$\lambda$"
                  & $k+{11\over 2}$
                  & $(0, \12)$ 
                  & $\32$ 
                  & $(0,k,1)$
                  & ${\bf 4}^*,{\bf 20}^*,{\bf 60}^*$
                \\ \hline
$\O _k ^{(14)}   \sim \tr F_+ \lambda \bar \lambda \bar \lambda X^k$ 
                  & $Q^3\bar Q^2$
                  & $\psi _{(\alpha)}$
                  & $k+{13\over 2}$
                  & $(\12, 0)$ 
                  & $\12$ 
                  & $(1,k,2)$
                  & ${\bf 36}^*,{\bf 140}^*,{\bf 360}^*$
                \\ \hline
$\O _k ^{(15)}   \sim \tr F_+ F_- \lambda X^k$ 
                  & $Q^3\bar Q^2$
                  & $\psi _\mu$
                  & $k+{11\over 2}$
                  & $(\12, 1)$ 
                  & $\12$ 
                  & $(1,k,0)$
                  & {\bf 4},{\bf 20},{\bf 60}
                \\ \hline
$\O _k ^{(16)}   \sim \tr F_+ F_- ^2 X^k$ 
                  & $Q^4\bar Q^2$
                  & $A_{\mu \nu}$
                  & $k+6$
                  & $(1, 0)$ 
                  & $1$ 
                  & $(0,k,0)$
                  & {\bf 1$_c$},{\bf 6$_c$},{\bf 20'$_c$}
                \\ \hline
$\O _k ^{(17)} \!  \sim  \!\tr \!  F_+ \! F_- \lambda \bar \lambda X^k$ 
                  & $Q^3\bar Q^3$
                  & $h_{\mu \alpha} \ \ a_{\mu \alpha \beta \gamma}$
                  & $k+7$
                  & $(\12, \12)$ 
                  & $0$ 
                  & $(1,k,1)$
                  & {\bf 15},{\bf 64},{\bf 175}
                \\ \hline
$\O _k ^{(18)}   \! \sim \! \tr F_+ ^2 \bar \lambda \bar \lambda X^k$ 
                  & $Q^4\bar Q^2$
                  & $A_{\alpha \beta}$
                  & $k+7$
                  & $(0, 0)$ 
                  & $1$ 
                  & $(0,k,2)$
                  & {\bf 10$_c$},{\bf 45$_c$},{\bf 126$_c$}
                \\ \hline
$\O _k ^{(19)}  \! \sim \! \tr F_+^2 F_- \bar \lambda    X^k$ 
                  & $Q^4\bar Q^3$
                  & $\psi _{(\alpha)}$
                  & $k+{15\over 2}$
                  & $(0, \12)$ 
                  & $\12$ 
                  & $(0,k,1)$
                  & ${\bf 4}^*,{\bf 20}^*,{\bf 60}^*$
                \\ \hline
$\O _k ^{(20)}  \! \sim \! \tr F_+ ^2 F_- ^2 X^k$ 
                  & $Q^4\bar Q^4$
                  & $h_\alpha ^\alpha \ \ a_{\alpha \beta \gamma \delta}$
                  & $k+8$
                  & $(0, 0)$ 
                  & $0$ 
                  & $(0,k,0)$
                  & {\bf 1},{\bf 6},{\bf 20'}
                \\ \hline
\end{tabular}
\end{center}
\caption{Super-Yang-Mills Operators, Supergravity Fields and $SO(4,2)\times U(1)_Y \times SU(4)$ Quantum Numbers. The range of $k$ is $k\geq 0$, unless otherwise specified. }
\label{table:1}
\end{table}
\subsection{The Use of \boldmath
$U(1)_Y$ Symmetry and  Anomalous Dimensions}

The automorphism group $U(1)_Y$ of the conformal $\N=4$ supersymmetry algebra
provides a very useful tool in the organization of the operators and
correlation functions in supergravity. $U_Y(1)$ transformations
are {\it not}  symmetries of the full IIB string theory, nor
of the ${\cal N}=4$ SYM theory. However this symmetry is recovered
in the supergravity limit, and hence the AdS/CFT duality predicts
that the SYM theory in the limit $N \to \infty$, $\lambda \to
\infty$
 is invariant under  $U_Y(1)$ \cite{intriligator}.

From the Table, it is clear that the
composite operator
\be
\O_B \equiv {1 \over \sqrt 2} \{  \ophi + i \oc\}
\ee
has maximal hypercharge $Y=2$ amongst the chiral operators.
In fact, it is the only single--trace operator with
this maximal hypercharge value.
By the same token, the operator $[\O_B \O_B]$ is the only
double trace operator with the maximal hypercharge of $Y=4$.
For example, the stress tensor has $Y=0$ and thus $[TT]$ has vanishing
$Y$ as well.

As supergravity has exact $U(1)_Y$ symmetry,
this guarantees that the operator $[\O_B \O_B]$ does not
mix in the limit of large $N$ and large $\lambda$
with any other operator at all, and that its anomalous
dimension can be read off from only the correlators already
computed above.
We must also expect
that the real and imaginary components of $[\O_B \O_B]$,
which are $[\ophi \ophi - \oc \oc]$ and $[\ophi \oc]$,
have the same anomalous
dimension.
This is indeed realized in the
above correlation functions.

The anomalous dimension of  $[\ophi \oc]$ can be immediately
read off from (\ref{phic}). To order $O(\frac{1}{N^2})$ and
for $\lambda \to \infty$
\bea
y \frac{d}{dy} \,[\ophi \oc]&  = &\gamma_{[\ophi \oc]}\; [\ophi \oc] \\
\gamma_{[\ophi \oc]}& =& \frac{A}{N^2}= 
-\frac{16}{N^2} \,.\nonumber
\eea
From (\ref{ope2phi}--\ref{Cphic}) we deduce that the action of
the dilation operator on the subspace of operators spanned
by  $[\ophi \ophi]$ and  $[\oc \oc]$ is
\bea
y{d \over\partial y}
\left(\begin{array}{cc}\big[ \ophi  \ophi\big ]\\\big[ \oc  \oc\big ]\end{array}\right)
=\frac{1}{N^2}
\left(\begin{array}{cc}C \quad B \\B \quad C \end{array}\right)\;
\left(\begin{array}{cc}\big[ \ophi  \ophi\big ]\\\big[ \oc  \oc\big ]\end{array}\right)\,.
\eea
As already explained, the space of operators of approximate dimension
8 is much bigger than this two--dimensional subspace, and we
do not have at present enough information
to fill the entries of the full mixing matrix.
However, the operator  $[\ophi \ophi - \oc \oc]$, which
is the eigenvector of this $2 \times 2$ matrix of eigenvalue
$(-B+C)/N^2$, has maximal $U_Y(1)$ charge $|Y|= 4$
and we can isolate its anomalous dimension
\be
\gamma _{[\ophi \ophi - \oc \oc]} = \frac{-B+C}{N^2} = -\frac{16}{N^2}\, ,
\ee
as expected. The fact that $A= -B+C$ as required by $U_Y(1)$
symmetry is a nice check on our calculation.


All other double--trace operators of approximate dimension 8
will have $|Y| < 4$. 
In the dilaton/axion sector of the theory,
we only encounter operators with $Y=0$, such as $[\O_B ^* \O_B]$
and $[T_{\mu \nu} T^{\mu \nu}]$. One particular linear combination of all
these $Y=0$ operators is known to be protected. This is the descendant
of the protected double trace operators
\be
Q ^4 \bar Q^4 \bigl \{ \tr X^2 \tr X^2 \} \big |_{{\bf 105}}
\ee
where in the tensor product of the two {\bf 20'} of $\tr X^2$,
only the representation of dimension 105 is retained.

\subsection{Comparison with Large N SYM Calculations}

\label{fieldtheory}
The prediction for the anomalous dimension of the operator $[\ob \ob]$
to order $1/N^2$ obtained from sugra holds for infinitely large value of the
't Hooft coupling $\lambda=g_{YM}^2 N$
on the $\N=4$ SYM side. As the only window
to date into large 't Hooft coupling is via the Maldacena conjecture,
we do not have any direct checks available for the values of the
anomalous dimensions or for the space--time dependence of the correlation
functions from SYM.
\begin{figure}
\begin{center} \PSbox{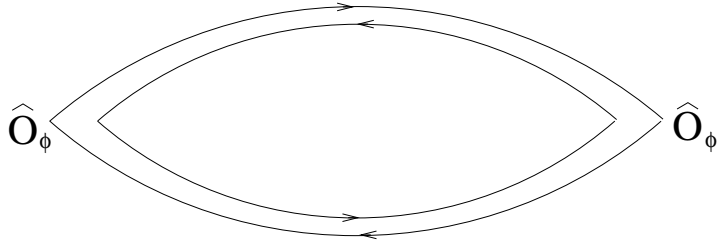 hscale=80 vscale=80}{3.5in}{1.2in} 
\end{center}
\caption{Born diagram for the
 2--point function $\< \hat \ophi \hat \ophi \>$ in the 
double--line representation.}
\end{figure}
However, it is very illuminating to reproduce the $1/N^2$ dependence
of the anomalous dimension from the standard large $N$ counting rules
of field theory, and to investigate any other consequences this may produce.
We proceed by expanding $\N=4$ SYM in $1/N$, while keeping the 't Hooft
coupling $\lambda$ fixed and perturbatively small. The strategy will be
to isolate the general structure of the expansion, then to seek the
 limit where $\lambda\to \infty$ and compare with supergravity
 predictions.

By way of example, we concentrate on the $\langle \ophi \oc \ophi\oc \rangle$
correlator, but the results apply generally. First, we normalize the
individual operators via their 2--point functions, as in (\ref{norm}).
To leading
order in $1/N$, this requires (up to numerical coefficients
we do not keep track of)
\bea
\hat \ophi & = & {1 \over N} \tr FF + \cdots
\\
\hat \oc & = & {1 \over N} \tr F\tilde F + \cdots
\eea
To Born approximation, this normalization is easily obtained by inspection
of Fig.5 : each operator has a $1/N$ normalization factor, and there are
two color loops, producing each a factor of $N$. According to general
non--renormalization results, the 2--point function is actually independent
of $g_{YM}$ and thus independent of the 't Hooft coupling.

With this normalization, the disconnected graph  in Fig.6 (a) contributes
precisely to order $N^0$. The simplest connected Born graph (b) of Fig.6
has a factor of $1/N^4$ from the normalizations of the 4 operators,
and two color loops, so its net contribution is of order $1/N^2$,
as expected. Finally, in graph (c) of Fig.6, we illustrate the higher
order perturbative contributions with a graph of order $g_{YM}^6$.
With a factor $1/N^4$ from operator normalization, and 5 color loops,
its total dependence is $g_{YM}^6 N =   \lambda^3/N^2$.
Thus, for fixed 't Hooft coupling, the $N$--dependence of graphs (b) and (c)
are the same, as expected. In fact, all planar graphs have
this same $N$--dependence to leading order in $N$ for fixed 't
Hooft coupling, and we
thus expect the connected part of the correlator to behave as
\be
\langle \hat \oc \hat \ophi \hat \oc \hat \ophi\rangle
\sim
{1 \over N^2} f(\lambda)
\ee
for some function $f$ of the 't Hooft coupling and position variables.

The above result was established perturbatively in the 't Hooft
coupling. To compare with supergravity
 results, $f$ ought to have a finite
limit as $\lambda \to \infty$. The Maldacena conjecture predicts
that it does, and gives a specific value for the limit.
It would be interesting to explicitly compute the anomalus dimension
of $[\ophi \oc]$ in perturbation theory. From the previous
discussion,
\be
\gamma_{[\ophi \oc]} =\frac{1}{N^2}(\gamma_1 \lambda
+ \gamma_2 \lambda^2 +\dots) + O\left(\frac{1}{N^4}\right)\,.
\ee
If the interpolation
between small and large $\lambda$ is a smooth cross--over,
it is natural to expect  the coefficient $\gamma_1$ to be negative,
as supergravity predicts a negative asymptotic value
for $\lambda \to \infty$.

While the $N$--dependence of the anomalous dimension of
 $[\ob \ob]$ follows
simply from large $N$ counting rules in SYM theory,
the space-time dependence of the correlators cannot be simply inferred
from SYM.
Supergravity
 results demonstrate that to order $1/N^2$, the 4--point correlator
has analytic behavior in position variables, except for a single
power of a logarithm. On the SYM side however, our
perturbative
treatment of the 't Hooft coupling prevents us from making any
sensible predictions on the space-time dependence of the correlators.
While graph (b) of Fig.6 is only power behaved, one expects graphs
like (c) to contain higher and higher powers of logarithms as larger and
larger numbers of virtual particles are being exchanged. The
Maldacena conjecture
predicts that somehow, as $\lambda \to \infty$, all these
powers and logarithms rearrange themselves and combine into a single
logarithm.
\begin{figure}
\begin{center} \PSbox{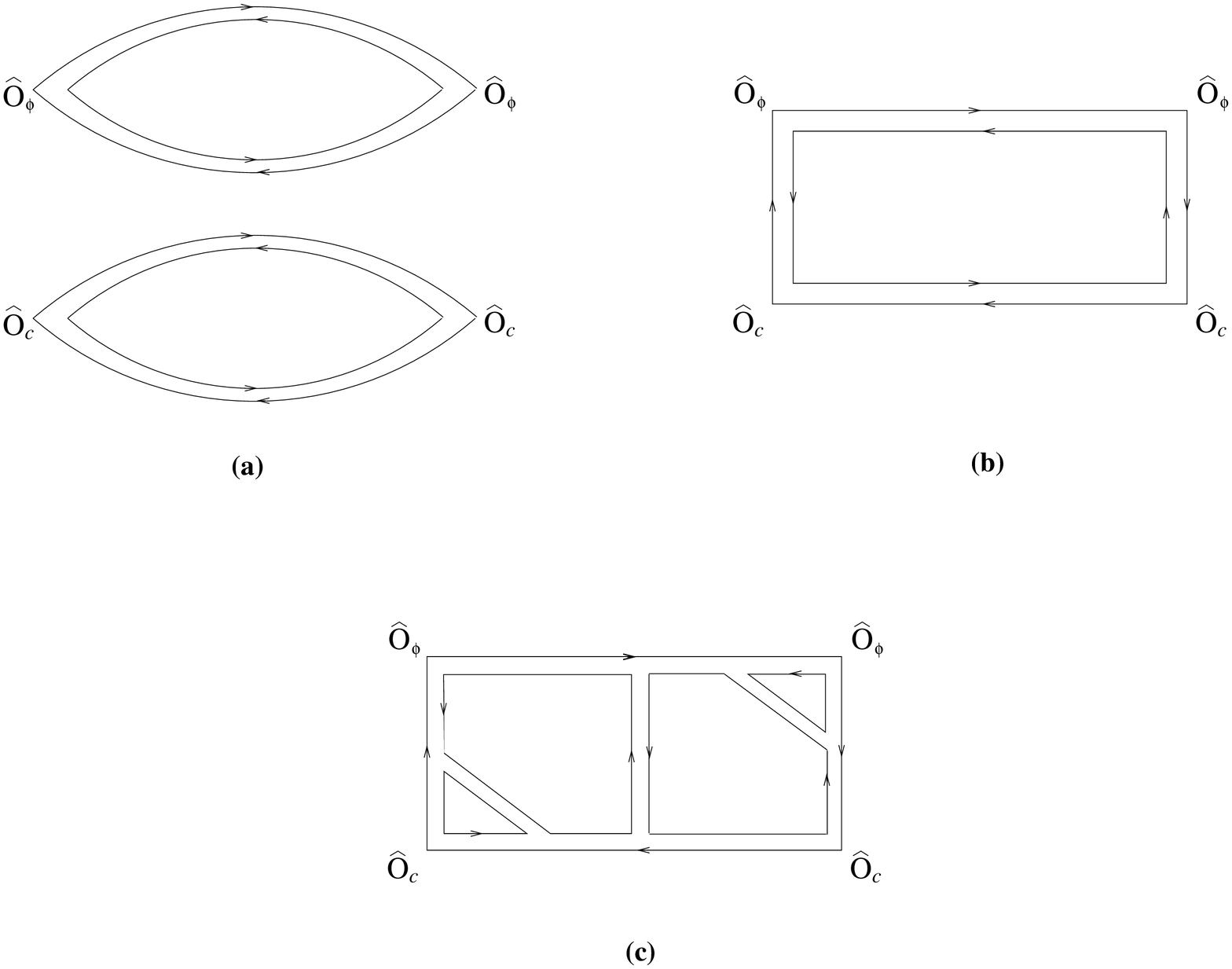 hscale=57 vscale=57}{6.5in}{3.5in} 
\end{center}
\caption{Some Feynman diagrams in the double--line representation
contributing to the 4--point function
$\<\hat \ophi \hat \oc \hat \ophi \hat \oc\>$.}
\end{figure}

\section{Conclusions}

We have shown that the supergravity results for 4--point functions
of chiral SYM operators can be successfully
interpreted in terms of a 4d Operator Product Expansion. There is
a generic relation between the power--singular terms that arise
in the direct channel limit of an  exchange diagram
and the contributions to 4--point function of the corresponding
`conformal partial wave'.
Logarithmic singularities can be naturally understood in terms
of $O(\frac{1}{N^2})$ renormalization and mixing
of the double--trace composites that arise in the OPE of two
single--trace chiral operators.

It should be emphasized that the very possibility of a 4d OPE interpretation
is quite non--trivial. For example, generic exchange integrals
in $AdS_5$ contain $(\log)^2$ singularities\footnote{This happens
when the exchange integral is equal to an infinite sum of quartic graphs. It
appears that all exchanges that arise in  IIB supergravity
on $AdS_5 \times S^5$ can be reduced to a
{\it finite} sum of quartics graphs \cite{howto}.}, which would be
impossible to interpret as $O(\frac{1}{N^2})$ renormalization
effects. It is then crucial that the couplings of IIB supergravity
on $AdS_5 \times S^5$ do not allow this type of processes.

We did not attempt to interpret the series expansions of the
supergravity 4--point correlators beyond the leading logarithmic
term.
It would be of great interest
to extend the analysis of this paper to the higher order terms.
The expansion of the supergravity amplitudes as the operator
insertions become close (see {\it e.g.} (\ref{Wexp})) is given
by series with a finite radius of convergence. This could be
regarded as an indication that the the 4d SYM theory admits
a convergent OPE.

A comment is in order about the issue of decoupling
of operators dual to string states. We were able to match
the power--singular terms and the leading log with the contributions
to the OPE of chiral primaries and their double--trace
products, so our results  support the idea
that at strong 't Hooft coupling
it is consistent to ignore the string states.
One would ultimately like to show that
for each possible limit as the boundary insertion points become
pairwise close, the
expansion of the 4--point supergravity amplitude
 can be interpreted to all orders as
a convergent double OPE in terms of the subset of operators
given by the chiral operators and their
multi--trace products. This would prove that as
$\lambda \rightarrow \infty$ this
subset  forms a closed algebra.

To tackle these issues it might be more convenient
to consider a 4--point function of ${\cal N}=4$ chiral primaries, the simplest
example being the correlator of four ${\bf 20'}$ operators
${\rm tr} X^{(i} X^{j)}$. The field theory analysis is somewhat cleaner
than in the dilaton/axion sector since  there are fewer double--trace
operators of approximate dimension 4 than of approximate
dimension 8. On the supergravity
side the computation involves exchanges of the ${\bf 20'}$, of the massless
vector and of the massless graviton,
as well as one quartic graph. All the exchange integrals
are available and the only missing piece of information is the numerical
value of the quartic
interaction vertex of four ${\bf 20'}$. As noticed in \cite{lee},
it should be possible to obtain this vertex directly from the 5d
gauged supergravity Lagrangian. The OPE of two  ${\bf 20'}$
contains double--trace operators in the ${\bf 105}$,  ${\bf 84}$,
 ${\bf 20'}$ and singlet representations. The  ${\bf 105}$ operator
is known to be protected, and it has been recently argued
\cite{ferrarashortening} that the   ${\bf 84}$ should also be protected
due to another shortening condition of the
${\cal N}=4$ superconformal algebra. The   ${\bf 20'}$ double--trace
 operator
is not protected but the superconformal algebra constrains
its anomalous dimension to be positive. Finally the singlet
is the `parent' chiral--primary operator from which the
$[{\cal O}_B{\cal O}_B]$ descendent
operator considered in this paper is obtained
by applying 8 Q's: hence its anomalous dimension must be the same
as $\gamma_{[{\cal O}_B{\cal O}_B]}$. It would be nice
to check these facts through an explicit supergravity calculation.

Finally, we would like to remark that although we have confined
our investigation to the 4d ${\cal N}=4$ SYM theory, our methods
apply to other AdS/CFT dualities in various dimensions.
In particular, the results
of Section 2 imply that for any CFT
that has an AdS dual,
in the limit in which the gravity approximation is valid
all the singular terms in the OPE of two protected operators are given
by other protected operators and their normal--ordered products.

\section*{Acknowledgments}
It is a pleasure to thank Dan Freedman for very enjoyable collaboration
in previous projects that have led to this investigation, and for
his advice in the present project as well.
We are grateful to Edward Witten for important discussions.
E.D'H. gratefully acknowledges the warm hospitality
offered by the Laboratoires de Physique Th\'eorique
at Ecole Polytechnique and at Ecole Normale Sup\'erieure,
as well as the support provided by the Centre National de Recherche
Scientifique (CNRS).

The research of E.D'H. is supported in part by  NSF Grant PHY-98-19686,
A.M. by  NSF grant PHY-9870115,
and L.R. by D.O.E. cooperative agreement
DE-FC02-94ER40818 and by INFN `Bruno Rossi' Fellowship.

\appendix

\setcounter{equation}{0}
\section{Summary of Supergravity 4--point Functions}

Exchange amplitudes for massive scalars, massive vectors and
massless gravitons with external scalar operators were evaluated in
\cite{dhfgauge, dhfscalar, march99, howto} in general
$AdS_{d+1}$ space.
 We summarize here the results
for massive scalar amplitudes with non--derivative couplings, and
for massless graviton exchange, defined by
\bea \label{A1}
I^{(t)}_{\rm scal} & =  &\int d z \sqrt g \int d w \sqrt g \
K_{\Delta_1}(z,x_1) K_{\Delta _3} (z,x_3)\
G_\Delta (z,w) \ K_{\Delta _2} (w,x_2) K_{\Delta _4} (w,x_4)
\\
I^{(t)}_{\rm grav} & = & \frac{1}{4} \,\int d z \sqrt g \int d w \sqrt g \
T_{13} ^{\mu \nu} (z) \
G_{\mu \nu\mu'\nu'} (z,w) \ T_{24} ^{\mu ' \nu'} (w)\, .
\label{A2}
\eea
Here the coupling of the graviton propagator is to
 to a conserved stress tensor,
given by
\bea \label{stress}
T_{13}^{\mu\nu}(z,  x_1,  x_3) &= & ~ D^{\mu} K_{\Delta_1} (z,  x_1)
 D^{\nu} K_{\Delta_1} (z,  x_3) \\
&& - \frac{1}{2}\,g^{\mu \nu} \bigl [ D_{\rho} K_{\Delta_1} (z, x_1)
D^{\rho} K_{\Delta_1}
(z, x_3)
+ m_1^2 K_{\Delta_1} (z, x_1) K_{\Delta_1} (z,  x_3) \bigr ]\, ,
 \nonumber
\eea
with an analogous expression for $T_{24}^{\mu'\nu'}$.
The superscript `$t$' in (\ref{A1}--\ref{A2}) indicates that
these integrals define what we call `t--channel' exchanges.
In our terminology, the  graphs in the s-- and u--
 channels are obtained from those in the
t--channel by  letting respectevely
$(x_1, x_2, x_3, x_4) \to (x_1,
x_3, x_2, x_4)$ and $  (x_1, x_2, x_3, x_4) \to (x_1, x_2, x_4, x_3)$.

Adopting the methods of \cite{howto}, the evaluation of these integrals
does not require the explicit form
of the bulk--to--bulk propagators, but only
their equations of motion:
\be \label{scalardelta}
( -\Box  + m^2)G_\Delta (u) = \delta(z,w)
\ee
\bea \label{gravitonpropequ}
&&-  D^\sigma D_\sigma G_{\mu \nu \mu ' \nu '}
  -  D_\mu D_\nu G_{\sigma} {}^\sigma {}_{\mu ' \nu '}
  +  D_\mu D^\sigma G_{ \sigma \nu  \mu ' \nu '}
 +  D_\nu D^\sigma G_{\mu \sigma  \mu ' \nu '}
 -2(G_{\mu \nu \mu ' \nu '} - g_{\mu \nu} G_\sigma {}^\sigma
{}_{\mu ' \nu'}) \nonumber \\
&& =
 \Bigl(g_{\mu \mu'}g_{\nu \nu'} +g_{\mu \nu'}g_{\nu \mu'} -  {2\over d-1}
g_{\mu\nu}g_{\mu' \nu'}\Bigr)
\delta(z,w)  + D_{\mu '} \Lambda _{\mu \nu \nu'} + D_{\nu '}\Lambda _{\mu \nu
\mu'} \nonumber
\eea
Remarkably, for all cases that arise in the IIB compactification
on $S^5$, the amplitudes can be expressed as simple linear combinations
of a finite number of 4--point quartic graphs.
It is
convenient to define
the integrals associated with these general quartic graphs as follows:
\be \label{D}
D_{\Delta _1 \Delta _3 \Delta _2 \Delta _4} (x_1, x_3, x_2, x_4)
=
\int {d^{d+1} z\over z_0 ^{d+1}}
\tilde K _{\Delta _1} (z,x_1)
\tilde K _{\Delta _3} (z,x_3)
\tilde K _{\Delta _2} (z,x_2)
\tilde K _{\Delta _4} (z,x_4)
\ee
Here, the bulk--to--boundary propagators  are given by \cite{witten}
\cite{april}
\be
K_\Delta (z,x) = C_\Delta \tilde K_\Delta (z,x)
= C_\Delta \biggl ( {z_0 \over z_0 ^2 + (\vec{z} - \vec{x})^2 } \biggr )
^\Delta
\ee
with the normalization
\be
C_\Delta
= {\Gamma (\Delta) \over \pi ^{\frac{d}{2}} \Gamma (\Delta - \frac{d}{2})}\,.
\ee
We introduce a special short--hand notation for some special
quartic graphs which will appear frequently\footnote{Up to a number of external $x_i$-dependent factors, these quantities equal the
function $W_\Delta ^{\Delta'}(0,0)$ and $W_\Delta ^{\Delta'}(1,0)$
introduced  in \cite{march99}.}
\bea
W_\Delta ^{\Delta '} (x_i)
& = & D_{\Delta \ \Delta ' \ \Delta \ \Delta'}(x_1, x_3, x_2, x_4)
\\
\bar W_\Delta ^{\Delta '} (x_i)
& = & D_{\Delta +2 \ \Delta ' \ \Delta  \ \Delta'}(x_1, x_3, x_2, x_4)
\eea
The expression for the exchange graphs are then as follows.
The massive scalar amplitude is
\bea \label{scalarD}
I^{(t)}_{\rm scal} &= & C_{\Delta_1} C_{\Delta_2} C_{\Delta_3} C_{\Delta_4}
\sum _{p=p_{\rm min}} ^{p_{\rm max}}
{N _p \over
(x_{13} ^2)^{\Delta _3 -p} }D_{\Delta _{13}+p\ p \ \Delta _2 \ \Delta _4}
(x_1, x_3, x_2, x_4)
\\
\label{coeff}
N_p& =&\frac{1}{4}
{\g{\half(\D{1}+\D{3}-\Delta)}\g{\half(\D{1}+\D{3}+\Delta-d)}\g{p}\g{p+\D{13
}}
\over \g{p+1-\half\Delta+\half\D{13}}\g{p+1-{d\over
2}+\half\Delta+\half\D{13}}\g{\D{1}}\g{\D{3}}}.
\eea
Here, $\Delta_{13} \equiv \Delta_1 - \Delta_3$, $p_{\rm min} = \half (\Delta - \Delta _{13})$, and $p_{\rm max} =
\Delta _3 -1$.
The massless graviton amplitude for $d=4$ is given by
\be
I^{(t)}_{\rm grav}  =  \Delta C_\Delta ^2 C_{\Delta '}^2
\biggl [({9\over 8}\Delta -1 -  \Delta ') W_\Delta ^{\Delta '}
+\sum _{k=1} ^{\Delta -1}
    M_k W_k ^{\Delta '}
    -{2 \Delta (\Delta ')^2 \over \Delta -1} \bar W^{\Delta '+1} _k
    - s {\Delta (\Delta ')^2 \over \Delta -1} W_k ^{\Delta '+1}
    \biggr ]
\ee
where the constants $M_k$ are defined by
\be
M_k = {\Delta (3\Delta -8)\over 8 (\Delta -1)^2} (-(\Delta ')^2 + 4 \Delta ' +3)
 + {\Delta \Delta ' (\Delta '-2) \over \Delta -1}\,.
\ee
For $d=4$, $\Delta=\Delta'=4$ we have:
\be \label{grav4}
I_{\rm grav} ^{(t)} =
-{2^5 \cdot 3^3  \over \pi ^8} \biggl [
15 W_4 ^4  +  \sum _{p=1} ^3 \bigl \{
-17 W_p ^4  +64 \bar W_p ^5
+ 32 s W_p ^5 \bigr \} \biggr ]\,.
\ee

\subsection{Explicit Form of Quartic Graphs and Series Expansion}

All 4--point functions depend non--trivially on two conformal
invariant cross ratios of the points $x_i$. We find it convenient to choose
the combinations $s$ and $t$ defined in (\ref{st}) of the text.
For Euclidean positions $x_i$, the ranges for these combinations are
$0\leq s \leq 1$ and $-1 \leq t \leq 1$.

It was shown in \cite{march99} that the 4--point quartic
functions $W_p ^q$ may be
expressed as follows
\be \label{WI}
W_p ^q (x_i)= {(-)^{p + q} \pi^{\frac{d}{2}} \Gamma (p +q -\frac{d}{2})
\over \Gamma (p)^2 \Gamma (q)^2 (x_{13}^2)^p (x_{24}^2)^q} \ s^q
\left ( {\partial \over \partial s }\right ) ^{q-1}
\bigg \{ s^{p-1} \left ( {\partial \over \partial s }\right ) ^{p-1} I(s,t)
\biggr \}
\ee
0where the universal function $I(s,t)$ is given by
\be
I(s,t) = \int _0 ^\infty d \mu \int _{-1} ^1 d \lambda
{1 \over \mu + s(1-\lambda^2)} \ {1 \over 1 +\mu + \lambda t}
\ee
The integral $I(s,t)$ is perfectly convergent and produces an analytic
function in $s$ and $t$, with logarithmic singularities in $s$ and $t$.

\newpage

\noindent
{\it (a) Direct channel series expansion}

The direct channel limit is given by $|x_{13}| \ll |x_{12}|$ or/and $|x_{24}|
\ll |x_{12}|$, so that $s,t \to 0$. We find
\bea \label{Wexp}
W_p^q(x_i) & \equiv & D_{p\,p\,q\,q}(x_i) =
  {(-)^{p+q} \pi ^{\frac{d}{2}} \Gamma (p+q -\frac{d}{2}) \over  \Gamma (p)^2
  \Gamma (q )^2 (x_{13}^2)^p (x_{24}^2)^q}
\sum _{k=0} ^\infty
  {\Gamma (k+1)^2 \ s^{k+1} \over \Gamma (k-p+2) \Gamma (k-q +2)}
\nonumber\\
&& \biggl \{b_k(t) - a_k(t) \bigl [ \ln s +2 \psi (k+1) - \psi (k-q +2)
-\psi (k-p+2)
\bigr ] \biggr \}
\eea
and
\bea \label{wbar}
\bar W_p^{q+1}(x_i) & = &
 =
  {(-)^{p+q} \pi ^{\frac{d}{2}} \Gamma (p+q-\frac{d}{2}+1)
  \over 4 \Gamma (p) \Gamma (p+2)
  \Gamma (q +1)^2 (x_{13}^2)^p (x_{24}^2)^q}
\sum _{k=0} ^\infty
  {\Gamma (k+1) \Gamma (k+3) \ s^{k+1}
  \over \Gamma (k-p+2) \Gamma (k-q +2)}\\
&&\cdot\biggl \{
 \hat b_k(t) - \hat a_k(t) \bigl [ \ln s + \psi (k+1) + \psi (k+3) - \psi
(k-q +2) -\psi (k-p+2)
\bigr ] \biggr \}
 \nonumber
\eea
Here the coefficient functions are given by
\bea
a_k(t) &
 = {\displaystyle \int _{-1} ^1 \! d\lambda {(1-\lambda ^2) ^k \over
(1+\lambda
t)^{k+1}}}
\qquad
{\displaystyle b_k(t)
 = \int _{-1} ^1 \! d\lambda {(1-\lambda ^2) ^k \over (1+\lambda t)^{k+1}}
  \ln {1+\lambda t \over 1-\lambda ^2}} &\label{(4.31a)}\\
{\displaystyle \hat a_k(t)} & {\displaystyle
 = \int _{-1} ^1 \! d\lambda {(1-\lambda ^2) ^{k+1} \over (1+\lambda
t)^{k+1}}}
\qquad
{\displaystyle \hat b_k(t)
 = \int _{-1} ^1 \! d\lambda {(1-\lambda ^2) ^{k+1} \over (1+\lambda
t)^{k+1}}
  \ln {1+\lambda t \over 1-\lambda ^2}} &\label{(4.31b)}
\eea
The coefficient functions admit Taylor series expansions in powers of $t$
with radius of convergence 1. In particular the functions
$a_k(t)$ have the following representation in terms
of hypergeometric series:
\be \label{ahyper}
a_k(t)=
\sqrt{\pi}{\Gamma(k+1) \over \Gamma\left(k+{3\over 2}\right)}\,
F\left({k+1\over 2},{k\over 2}+1;k+{3\over 2};t^2\right).
\ee
We have the  following relations between these
functions
\bea
(k+2) \hat a_k (t)\: &
 =& (k+1) \bigl (2a_k(t) - a_{k+1}(t) \bigr )
  \label{(4.32a)} \\
(k+2)^2 \:\hat b_k (t) &
 =& (k+1)(k+2) \bigl (2b_k(t) - b_{k+1}(t) \bigr ) -2a_k(t) + a_{k+1}(t)
  \label{(4.32b)}
\eea
The presentation of these series expansions is slightly formal in the sense
that for $k\leq q -2$, the $\Gamma (k-q +2)$ function in the denominator
produces a zero, while the $\psi (k-q +2)$ term produces a pole, which
together yield a finite result, which amounts to a pole term in $s$.
Its coefficient can be obtained from the formula
$ \lim_{x \rightarrow 0} \psi(x-q)/ \Gamma(x-q) = (-)^{q+1} \Gamma(q+1)$
for any non--negative interger $q$.

\noindent

\newpage

{\it (b) Crossed channel series expansion}

The crossed channel asymptotics is given by $s \to \half$ and $t \to \pm 1$.
The `s--channel' limit  $|x_{12}| \ll |x_{13}|$ or/and
 $|x_{34}| \ll |x_{13}|$ corresponds to $s \rightarrow 1/2$,
$t \rightarrow -1$, while the `u--channel' limit $|x_{23}| \ll |x_{34}|$
and/or $|x_{14}| \ll |x_{34}|$ corresponds to $s \rightarrow 1/2$, $t
\rightarrow +1$.

Both limits were derived in \cite{march99}
 by first obtaining a suitable expansion for
the universal function $I(s,t)$ and then using (\ref{WI}).
This expansion may be used
to evaluate the logarithmic part of $W_p^q$ and we obtain
\bea \label{Wcrossedexp}
W_p ^q(x_i)\bigg |_{{\rm log}} & =&
 -  \ln (1-t^2)
{2^{p-2} \pi ^{\frac{d}{2}} \Gamma (p +q -\frac{d}{2}) \over
 \Gamma (p) \Gamma (q) (x_{13}^2)^p (x_{24}^2)^q}\nonumber\\
&& \sum _{\ell =0} ^{q-1}
\sum _{k=0}^\infty
{(-2)^{-\ell} \Gamma(k+1) \ s^{p +q-\ell-1} \ (1-2s)^{k-p+\ell-q+2}
\over
\Gamma (q-\ell) \Gamma (p-\ell) \ \ell ! \ \Gamma (k+\ell -p-q+3)}
\ \alpha _k(t)
 \label{(4.40)}
\eea
Notice that in the crossed channel, no power singularities arise.
The coefficient functions $\alpha _k(t)$ are given by
\be \label{alpha}
\alpha _k(t) = \sum _{\ell =0} ^\infty
{\Gamma (\ell + \half) \over \Gamma (\half) \ell !}
{(1-t^2) ^\ell \over 2 \ell + k +1} \,.
\ee

\setcounter{equation}{0}
\section{Matching Supergravity Exchanges and  Partial Waves}

\subsection{Conformal Partial Amplitude}
Here we present the partial amplitude
$\<\O_1(x_1)\O_2(x_2)\O_3(x_3)\O_4(x_4)\>_0$, which corresponds to the
contribution to the 4--point function of a scalar operator $\O_\Delta$
from the OPE of $\O_1(x_1)\,\O_3(x_3)\sim \O_\Delta(x_1)/
x_{13}^{\D{1}+\D{2}-\Delta}$
and all its conformal descendants.
In a CFT, an expression for the partial amplitude can be written as
\be
\<\O_1(x_1)\O_2(x_2)\O_3(x_3)\O_4(x_4)\>_0=
{1\over x_{13}^{\S{13}-\D{13}} x_{12}^{\D{13}+\D{24}} x_{14}^{\D{13}-\D{24}}
x_{24}^{\S{24}-\D{13}}}\;f_0(\rho,\eta)
\ee
where $\S{ij} \equiv \Delta_i + \Delta_j$, $\D{ij} \equiv \Delta_i -\Delta_j$
and
\be
\eta={x_{12}^2\, x_{34}^2\over x_{13}^2\, x_{24}^2},\;\;\;\;
\rho={x_{14}^2\, x_{23}^2\over x_{13}^2\, x_{24}^2}
\ee
are the two cross--ratios. The function $f_0(\rho,\eta)$ of the cross--ratios
was
obtained\footnote[1]{With normalizations such that
$\<\O_\Delta\O_\Delta\>,\;\<\O_1\O_3\O_\Delta\>,\;\<\O_1\O_3\O_\Delta\>$ are
all given by coefficient 1
times the appropriate conformally invariant function
of coordinates. This implies that in the OPE of
$\O_1\O_2$ the coefficient $C_{13\Delta}$ of the operator $O_{\D{}}$ is 1.}
in \cite{ferraraope} (equ.(3.2)):

\newpage

\bea
f_0(\rho,\eta)&=&\eta^{{1\over 2}(\D{13}+\D{24})} {\G (\Delta)\over
\G\left({1\over 2}(\Delta+\D{13})\right)
\G\left({1\over 2}(\Delta-\D{13})\right)}\:\\
&\times&\int_0^1\,d\sigma\,
\sigma^{{1\over 2}(\D{13}-\D{24})-1}\,
(1-\sigma)^{-{1\over
2}(\D{13}+\D{24})-1}\,
\left({\rho\over\sigma}+{\eta\over
(1-\sigma)}\right)^{-{1\over 2}(\D{24}+\Delta)}\nonumber\\&\times&
F\left(\half (\Delta-\D{24}),\half (\Delta+\D{24});\Delta+1-{d \over
2};\left({\rho\over\sigma}+{\eta\over
(1-\sigma)}\right)^{-1} \right)\nonumber
\eea
Our goal now is to rewrite this partial amplitude in the form of an
expansion in conformally invariant variables $s=\half {1\over \eta +
\rho}$ and $t= {\eta - \rho \over \eta + \rho}$ (see (\ref{st})), which will
allow a more direct comparison with the corresponding supergravity
exchange
diagram.
To do this, following the steps of Appendix A of \cite{ferraraope}
((A.1) to (A.3)), we first expand the hypergeometric function in a power series
\bea
&&F\left(\half (\Delta-\D{24}),\half (\Delta+\D{24});\Delta+1-{d \over
2};\left({\rho\over\sigma}+{\eta\over
(1-\sigma)}\right)^{-1} \right)=
\nonumber\\
&&\sum_{n=0}^\infty {1\over n!} {\left( \half (\Delta-\D{24})\right)_n\,\left(
\half (\Delta+\D{24})\right)_n\over
\left( \Delta+1-{d \over 2}\right)_n}\:\left({\rho(1-\sigma z)\over
\sigma (1-\sigma)} \right)^{-n},
\eea
where $z=1-{\eta\over\rho}.$
We then perform the integral in $\sigma,$ which gives
\bea
&&\int_0^1\,d\sigma\,
\sigma^{{1\over 2}(\D{13}+\Delta)+n-1}\,
(1-\sigma)^{-{1\over
2}(\D{13}+\Delta)+n-1}\,(1-\sigma
z)^{-\half(\D{24}+\Delta)-n}=\\&&{\textstyle
{\Gamma\left(\half(\D{13}+\Delta)+n\right)\,\Gamma\left(\half
(-\D{13}+\Delta)+n\right)
\over \Gamma\left(\Delta+2n\right)}\: F\left(\half
(\Delta+\D{13})+n,\half (\Delta+\D{24})+n;\Delta+2n;z \right)}. \nonumber
\eea
Putting everything together,
we get the expression for $f_0(\rho,\eta):$
\bea
f_0(\rho,\eta)&=&{\Gamma(\Delta)\eta^{\half (\D{13}+\D{24})}\over
\rho^{\half (\Delta+\D{24})}}\:\sum_{n=0}^\infty
{\left({\Delta+\D{13}\over 2}\right)_n\left({\Delta-\D{13}\over
2}\right)_n\left({\Delta+\D{24}\over
2}\right)_n\left({\Delta-\D{24}\over 2}\right)_n\over
\Gamma(\Delta+2n)\,(\Delta+1-{d\over 2})_n\, n!}\nonumber\\
&\times& {1\over\rho^n}\:
F\left(\half(\Delta+\D{13})+n,\half(\Delta+\D{24})+n;\, \Delta+2n;\,z\right)
\eea
We now restrict ourselves to a case of pairwise equal dimensions,
$\D{13}=\D{24}=0.$
We also change to $s,t$ variables:
\be
z= 1-{\eta\over\rho}={2t\over t-1},\;\;\;\; {1\over \rho}={4s\over
1-t}
\ee
Using quadratic transformation of hypergeometric function
\be
{F\left({\Delta\over 2}+n,{\Delta\over 2}+n;\,\Delta+2n;\,{2t\over
t-1}\right)\over (1-t)^{{\Delta\over 2}+n}}=
F\left({\Delta\over 4}+{n\over 2},{\Delta\over 4}+{n\over
2}+\half;\,{\Delta\over 2}+n+\half;\,t^2\right)
\ee
and the definition of $a_k(t)$ from (\ref{ahyper})
we can finally rewrite the partial amplitude as an expansion in
$s$ and $t$ variables as
\be
\<\O_1(x_1)\O_2(x_2)\O_3(x_3)\O_4(x_4)\>_0=
{1\over x_{13}^{2\D{1}}x_{24}^{2\D{2}}}\;g_0(s,t);
\ee
\be\label{partialfinal}
g_0(s,t)=2{\Gamma\left(\Delta\right)\Gamma\left(\Delta+1-{d\over
2}\right)\over
\Gamma\left({\Delta\over 2}\right)^4}\,
\sum_{n=0}^\infty\;{{\Gamma\left({\Delta\over
2}+n\right)^2\over \Gamma\left(\Delta+1-{d\over
2}+n\right)\,n!}\:s^{\half\Delta+n}\:a_{n+{\Delta\over 2}-1}(t)}.
\ee
Note that singular terms in the limit
$|x_{13}|\ll|x_{12}|$ correspond to
$n\leq\D{1}-\hd-1.$

\subsection{Singular terms from Witten diagrams}

In this subsection we will find the singular terms of a given
$AdS$ scalar exchange diagram in the form of an expansion in $s$ and $t$,
and comparing with the singular terms of the corresponding partial
amplitude from (\ref{partialfinal}) we will find an exact match.
We recall the result (\ref{scalarD}) that any exchange supergravity
diagram reduces to a sum of quartic graphs.
For simplicity, we restrict ourselves to the case of pairwise equal
dimensions, as in the analysis of partial amplitude in the previous
subsection, $\D{13}=\D{24}=0.$ We also recall the expansion of a
quartic graph  (\ref{Wexp}).
Let us assume that $\D{1}\leq\D{2}$, which means that $p_{\rm
max}\leq\D{2}-1.$ Upon insertion of the expansion (\ref{Wexp}) into
(\ref{scalarD}) we notice that the only power--singular terms
$O(\frac{1}{x_{13}^n})$ in the limit
$x_{13}\ll x_{12}$ are
the ones for which $k\leq\D{1}-2$. Keeping terms $k \leq \D{2} -2$
would amount to consider also all the power singularities
$O(\frac{1}{x_{24}^n})$ in the limit $x_{24} \ll x_{12}$. However
as will be apparent below only the terms that obey the
more stringent restriction
$k\leq\D{1}-2$ can be matched with the partial amplitude.
Using the relation
\be
{\psi(k-\D{2}+2) \over \g{k-\D{2}+2}}=(-1)^{k-\D{2}+1}\g{\D{2}-k-1}
\ee
we can now write explicitly the contribution of the singular terms
$O(\frac{1}{x_{13}^n})$ to
each quartic graph (notice that $p \leq \Delta_1-1$):
\be\label{sing}
D_{p\,p\,\D{2}\D{2}}\bigg |_{\rm sing}={(-)^{p} \pi ^\d \Gamma (p +\D{2} -\d)
\over  \g{p}^2
\g{\D{2}}^2{\left(x_{13}^2\right)^p \left(x_{24}^2\right)^{\D{2}} } }
\sum _{k=p-1} ^{\D{1}-2} {\g{k+1}^2\,s^{k+1}\over
\g{k-p+2}}a_k(t)\:(-)^{k+1}\,\g{\D{2}-1-k}
\ee
We are now in a position to extract power singularities from a scalar
exchange (\ref{scalarD}). Inserting (\ref{sing}) into
(\ref{scalarD}), we get

\newpage

\bea\label{doublesum}
&&\<\O_{\D{1}}\O_{\D{2}}\O_{\D{3}}\O_{\D{4}}\>\bigg |_{\rm
sing}=\prod_{i=1}^4C(\D{i})\,{\pi^\d
\over \g{\D{2}}^2}{1\over \left(x_{13}^2\right)^{\D{1}}
\left(x_{24}^2\right)^{\D{2}}} \times\nonumber\\
&&\sum_{k=0}^{\D{1}-2}\,\g{k+1}^2s^{k+1}(-)^{k+1}\,a_k(t)
\g{\D{2}-1-k}\:\sum_{p={\Delta\over
2}}^{\D{1}-1}\,N_p{(-)^p\g{p+\D{2}-\d}\over \g{p}^2\g{k-p+2}}
\eea
We will now evaluate the second summation in this formula. Due to the
pole of the gamma function in the denominator for $k-p+2 \leq 0$, one
can extend the $p$--summation to infinity. Notice that this would not
be possible if the $k$ summation  extended up to
$\Delta_2 -2$ instead of $\Delta_1-2$.
This gives
\bea
&&\sum_{p={\Delta\over
2}}^{\infty}\,N_p\,{(-)^p\g{p+\D{2}-\d}\over \g{p}^2\g{k-p+2}}=\\
&&{1\over
4}(-)^{k+1}{\g{\hd+\D{1}-\d}\g{\hd+\D{2}-\d}\g{\D{1}-\hd}\g{\D{2}-\hd}\over
\g{\D{1}}^2 \g{k+\hd-\d+2}\g{k+2-\hd}\g{\D{2}-1-k}}\nonumber
\eea
Putting this back into the expression for the scalar exchange
(\ref{doublesum})
and shifting the remaining $k$--summation to $n=k-\hd+1$ we finally
get
\bea\label{nonorm}
\<\O_{\D{1}}\O_{\D{2}}\O_{\D{3}}\O_{\D{4}}\>\bigg |_{\rm
sing}&=&{1\over 4\pi^{3d \over 2}}{1
\over \left(x_{13}^2\right)^{\D{1}}
 \left(x_{24}^2\right)^{\D{2}}} {\textstyle
{\g{\hd+\D{1}-\d}\g{\hd+\D{2}-\d}\g{\D{1}-\hd}\g{\D{2}-\hd}\over
\g{\D{1}-\d}^2\g{\D{2}-\d}^2
}}\nonumber \\&\times&
\sum_{n=0}^{\D{1}-\hd-1}s^{\hd+n}\,a_{\hd+n-1}(t)
{\g{\hd+n}^2\over\g{n+\Delta-\d+1}\,n!}.
\eea
Now, as in
(\ref{partialfinal}), we normalize
$\<\O_\Delta\O_\Delta\>,\;\<\O_1\O_3\O_\Delta\>,\;\<\O_2\O_4\O_\Delta\>$ to 1.
This can be achieved by noting
that the OPE coefficient $\O_i \O_j= C_{ijk}{O_k\over
x_{ij}^{\D{i}+\D{j}-\D{k}}}+\dots$ is
\be
C_{ijk}={{\cal A}_{\<\O_i\O_j\O_k\>}\over {\cal A}_{\<\O_k\O_k\>}},
\ee
where by ${\cal A}$ we denote the normalization of the corresponding
correlator.
To make $C_{ij\Delta}=1$ in the double OPE
of the 4--point function we multiply
(\ref{nonorm}) by ${{\cal A}_{\<\O_1\O_3\O_\Delta\>}\over
 {\cal A}_{\<\O_\Delta\O_\Delta\>}}\times
{{\cal A}_{\<\O_2\O_4\O_\Delta\>}\over {\cal A}_{\<\O_\Delta\O_\Delta\>}}$ and
then we further
divide by the normalization of $\<\O_\Delta\O_\Delta\>$ to normalize this
2--point function in the double OPE to
1. The appropriate coefficients,
\bea
{\cal A}_{\<\O_i\O_j\O_k\>}&
=& \textstyle{-\frac{
\Gamma[\frac{1}{2} (\Delta_i + \Delta_j - \Delta_k)]
\Gamma[\frac{1}{2} (\Delta_j + \Delta_k - \Delta_i)]
\Gamma[\frac{1}{2} (\Delta_k + \Delta_i - \Delta_j)]
 \Gamma[\frac{1}{2} (\Delta_i + \Delta_j + \Delta_k
-d) ] }{
2 \pi^d
\Gamma[\Delta_i - \frac{d}{2}]
\Gamma[\Delta_j - \frac{d}{2}]
\Gamma[\Delta_k - \frac{d}{2}]
}}\nonumber\\
{\cal A}_{\<\O_i\O_i\>}&=&\textstyle{2{\g{\D{}}(\D{}-\d)\over \pi^\d
\g{\D{}-\d}}}
\eea
can be found in \cite{april}.
We now arrive to the expression
\bea
\<\O_{\D{1}}\O_{\D{2}}\O_{\D{3}}\O_{\D{4}}\>\bigg |_{\rm sing}&=&{2
\over \left(x_{13}^2\right)^{\D{1}}
 \left(x_{24}^2\right)^{\D{2}}}
{\Gamma\left(\Delta\right)\Gamma\left(\Delta+1-{d\over
2}\right)\over
\Gamma\left({\Delta\over 2}\right)^4}\,\nonumber\\&\times&
\sum_{n=0}^{\Delta_1-\frac{\Delta}{2}-1}\;{{\Gamma\left({\Delta\over
2}+n\right)^2\over \Gamma\left(\Delta+1-{d\over
2}+n\right)\,n!}\:s^{\half\Delta+n}\:a_{n+{\Delta\over 2}-1}(t)}
\eea
which directly matches the singular terms
$O(\frac{1}{x_{13}^n})$ in (\ref{partialfinal}).

\setcounter{equation}{0}
\section{ Dilaton--Axion Correlators}

Let $\ophi$ and $\oc$ be the operators that couple to the dilaton and
axion supergravity fields with unit strength:
\be
S_{\rm{int}} = \int d^4 x \; \phi(x) \ophi(x) + C(x)  \oc(x) \,.
\ee
The kinetic terms in the 5--dimensional supergravity
action are normalized as
\be
\frac{1}{2 \kappa_5^2} \int_{AdS_5} \frac{1}{2} (\partial \phi)^2
+ \frac{1}{2} (\partial C)^2
\ee
where the gravitational coupling is given in terms of SYM parameters
(setting the AdS radius $R \equiv 1$) by
\be
2 \kappa_5^2 = \frac{8 \pi^2}{N^2} \,.
\ee
The 2--point functions are then (equ.(A.13) of \cite{april})
\be
\<\ophi(x_1) \ophi(x_2) \> =\<\oc(x_1) \oc(x_2) \>=
\frac{4 \cdot C_4 }{ 2\kappa_5^2}\frac{1}{x_{12}^8}
= \frac{3 N^2}{ \pi^4} \frac{1}{x_{12}^8}\,.
\ee
It is convenient to define normalized operators
$\phat = \zeta \ophi$, $\chat = \zeta \oc$, such that
\be \label{norm}
\<\phat (x_1) \phat(x_2)\> = \<\chat (x_1) \chat(x_2)\>=
\frac{1}{x_{12}^8}\,.
\ee
The normalization constant $\zeta$ is then
\be
\zeta =  {\pi \kappa_5 \over 2\sqrt 3 } =\frac{\pi^2}{\sqrt{3}N } \,.
\ee
The complete 4--point functions in the dilaton--axion sector
were assembled in \cite{march99}. Since the supergravity action is even under sign reversal of the axion
field, 3 different amplitudes enter: $\langle \hat \ophi \hat \ophi \hat \ophi \hat \ophi \rangle$,
$\langle  \hat \ophi \hat \oc \hat \ophi \hat \oc \rangle$ and
$\langle \hat \oc \hat \oc \hat \oc \hat \oc \rangle $.
Only graviton exchanges contribute to the
4--dilaton amplitude. The 4--axion amplitude is given by
graviton and dilaton exchanges, but it was checked in \cite{august}
that the dilaton contributions  precisely cancel when the
s, t and u channels are added together, so that
\be \label{4dil}
\langle \hat \ophi \hat \ophi \hat \ophi \hat \ophi \rangle
=\langle \hat \oc \hat \oc \hat \oc \hat \oc \rangle
 = I_0 + (2 \kappa_5^2)^{-1}\zeta ^4 \bigl
\{I_{\rm grav} ^{(s)} + I_{\rm
grav} ^{(t)} +I_{\rm grav} ^{(u)} \bigr \}\,.
\ee
The equality of the 4--dilaton and 4--axion amplitudes actually follows
directly from  the $U(1)_Y$ symmetry
of supergravity. Here, $I_0$ is the contribution
from disconnected graphs
and  $I_{\rm grav}^{(i)}$ is
the graviton exchange integral in the
channel $i$. The t--channel graviton exchange has been
given in terms of quartic graphs in (\ref{grav4}), the other channels
are readily obtained by permuting the coordinates $x_i$.
The disconnected contributions are easily
evaluated recalling the normalization (\ref{norm}):
\be \label{I0}
I_0 = \frac{1}{x_{12}^8 x_{34}^8}+ \frac{1}{x_{13}^8 x_{24}^8}+
\frac{1}{x_{14}^8 x_{23}^8}\,.
\ee
The constant $(2 \kappa_5^2)^{-1} \zeta ^4=
\frac{\pi^6}{72}\frac{1}{N^2}$ in front of the
exchange graphs in (\ref{4dil}) arises from a factor $2 \kappa_5^2$
for each bulk--to--bulk propagator, a factor $(2 \kappa_5^2)^{-1}$
for each vertex and from the normalization $\zeta$
of the dilaton and axion operators. As expected, the connected
graphs are $O(\frac{1}{N^2})$ relative to the disconnected part.

The amplitude $\langle \hat \ophi(x_1) \hat \oc(x_2) \hat \ophi(x_3) \hat
\oc(x_4) \rangle$
 is given by the t--channel graviton exchange, as well as axion exchanges and
a quartic interaction.
It was shown in \cite{august} that the sum of all the graphs
except the graviton exchange can be conveniently expressed as an
`effective' quartic graph
\bea
I_{q}^{\rm eff}
& = &
- 2 \int {d^5 z \over z_0 ^5} K_4(z,x_1) D_\mu K_4(z,x_2)
K_4(z,x_3) D^\mu K_4(z,x_4) \\
& = &
-{2^9 \cdot 3^4 \over \pi ^8} \bigl [W_4^4(x_i) - 4 s W_4 ^5 (x_i)\bigl ]
\eea
We then have
\be \label{fulldilax}
\langle \hat \ophi \hat \oc \hat \ophi \hat \oc \rangle
=  I'_0  + (2 \kappa_5^2)^{-1} \zeta ^4 \bigl \{ I_{\rm grav}^{(t)}
+ I^{\rm eff}_q \bigr \}\,.
\ee
Here the disconnected contribution $I'_0$ is simply
\be \label{I0p}
I'_0= \frac{1}{x_{13}^8 x_{24}^8} \,.
\ee

Notice that the 4--point functions
of axions and dilatons do {\it not} receive contributions
from  exchanges of
the `fixed scalar' of $m^2=32$ that corresponds \cite{vannieuv} to the
dilation mode of the $S^5$. The subtleties associated
with the computation of 3--point functions of two dilatons (or axions)
and a fixed scalar were recently discussed in 
\cite{liutseytlinfixedscalar, extremal}. Although naively
the  cubic
coupling of 2 dilatons and a fixed scalar is absent in the 5--dimensional
supergravity Lagrangian, the corresponding 3--point function can be
shown to be non--vanishing, either through a procedure of analytic
continuation in the KK level of the 
dilatons \cite{liutseytlinfixedscalar}, or by a careful
dimensional reduction that takes into account the contraints
of the supergravity fields and leads to non--vanishing
boundary interactions \cite{extremal}. 
The two procedures give identical
results \cite{extremal}. 
In the analytic continuation method,
one takes the conformal dimension of the dilatons to be $4+\epsilon$.
The cubic coupling dilaton--dilaton--fixed scalar vanishes as
$O(\epsilon)$ in the limit $\epsilon \to 0$, but 
the 3--point integral diverges as $O(\frac{1}{\epsilon})$, so that the
product gives a finite contribution.
 The procedure of analytic continuation
can be immediately used to prove that  fixed scalar exchanges give 
no contribution to the 4--point functions of dilatons and axions. In fact,
contrary to the 3--point function case, the exchange integral
with 4 external dimensions $4+\epsilon$ and bulk dimension 8
is  perfectly convergent in the limit $\epsilon \to 0$
(see {\it e.g.} \cite{dhfscalar}), 
whereas the cubic couplings vanish, yielding zero contribution.

\subsection{Leading Logarithm Asymptotics}
We now determine the leading logarithmic terms in the dilaton--axion
4--point functions
in the limits as the points become pairwise close. We adopt the
terminology:
\begin{quote}
a) t--channel limit: $|x_{13}| \ll |x_{12}|$, $|x_{24}| \ll |x_{12}|$,
which corresponds to $s,t \rightarrow 0$;\\
b) s--channel  limit:  $|x_{12}| \ll |x_{13}|$,
 $|x_{34}| \ll |x_{13}|$, or $s \rightarrow 1/2$,
$t \rightarrow -1$; \\
c) u--channel limit:  $|x_{23}| \ll |x_{34}|$,
  $|x_{14}| \ll |x_{34}|$, or  $s \rightarrow 1/2$, $t \rightarrow 1$.
\end{quote}

The limit of $I_{\rm grav} ^{(t)}$
for
  $s,t\to 0$ (t--channel)
was obtained in \cite{march99};
its logarithmic parts are given by
\bea
I_{\rm grav}^{(t)} \bigg |_{\rm log}  & =& {3\cdot 2^{3} \over \pi ^6} {\ln s
\over
x_{13}^8x_{24}^8}
\sum _{k=0} ^\infty s^{4+k} \ {\Gamma (k+4) \over \Gamma (k+1)} \bigg \{
-2(5k^2 +20k+16)(3k^2+15k+22) a_{k+3}(t) \nonumber\\&&
 + (k+4)^2 (15k^2 +55 k^2 +42) a_{k+4}(t) \bigg \}\,.
\eea
The analogous result for $I_{q}^{\rm eff}$ is
\be
 I_{q}^{\rm eff}\bigg |_{\rm log}
={2^6 \cdot 3 \cdot 5 \over
\pi^6}{\ln s\over  x_{13}^8 x_{24}^8}\sum_{k=0}^\infty
s^{k+4}\left\{(k+1)^2(k+2)^2(k+3)^2(3k+4)\:a_{k+3}(t)\right\}\,.
\ee
For our purposes, we shall need only the leading logarithmic contributions of
the amplitudes in the various channels. As we shall have to use permutations
on the points $x_i$ to find the exchange amplitudes in all channels, it will
be most convenient to re-express the $s$-- and $t$--dependence in terms of the
$x_i$ variables. Using the numerical values $a_3(0) = \frac{32}{35}$
and $a_4(0)=
\frac{256}{315}$, we find
\bea
I_{\rm grav}^{(t)}
& \sim &
-{ 2^7 \cdot 3 \over 7 \pi ^6 (x_{12}^{16}}
\ln \frac{x_{13}^2 x_{24}^2 }{x_{12}^4}
\qquad {\rm t-channel}
\\
I_q^{\rm eff}
& \sim &
+{2^7 \cdot 3^3 \over 7 \pi ^6 x_{12}^{16}} \ln \frac{x_{13}^2 x_{24}^2}
{x_{12}^4}
\qquad {\rm t-channel}
\eea

The limits in the crossed channels, where $s\to \half$ and $t^2 \to 1$, were
not obtained explicitly in \cite{march99}.
We shall now derive them here, by deriving the
crossed channel limits for the $W$-functions, starting from (\ref{Wcrossedexp}).
First, we notice from the definition of $\alpha _k(t)$ that $\alpha _k (\pm
1) = 1/(k+1)$. Next, it is clear from (\ref{Wcrossedexp}) that a non-trivial
leading logarithmic contribution will arise only if the orders of summation
satisfy $k=p+q-2-\ell \geq 0$. Since in our expressions $p,q\geq 1$ in all
cases, and $0\leq \ell \leq q-1$, the inequality above can always be realized
and there is always a $k\geq 0$ satisfying $k=p+q-2-\ell $. The remaining
summation over $\ell$ is then
\be
W_p ^q (x_i) \sim - \ln (1-t^2)
{\pi ^2 \Gamma (p+q-2) \over 2\Gamma (p) \Gamma (q) (x_{13}^2)^4 (x_{24}^2)^4}
\sum _{\ell =0} ^{q-1} (-)^\ell
{\Gamma (p+q-1-\ell) \Gamma (p+q-1-\ell)
\over
\Gamma (q-\ell) \Gamma (p-\ell) \Gamma (p+q-\ell) \ \ell !}
\ee
But this $\ell$-sum is precisely proportional to a hypergeometric function
${}_3F_2(p,p,1-q;p-q+1,p+1;1)$ evaluated at unit argument. As a result, we have
\bea
W_p ^q (x_i) & \sim & - {\pi ^2 \over 2} {1 \over (p+q-1) (p+q-2)}
{\ln (x_{12}^2 x_{34}^2/x_{13}^4) \over (x_{13}^2)^8 }
\qquad {\rm s-channel} \qquad
\\
W_p ^q (x_i) & \sim & - {\pi ^2 \over 2} {1 \over (p+q-1) (p+q-2)}
{\ln (x_{14}^2 x_{23}^2/x_{12}^4) \over (x_{12}^2)^8 }
\qquad {\rm u-channel} \qquad
\eea
Finally, we need the asymptotic behavior of the function $\bar W$ in this
channel with $s \to \half$ and $t^2 \to 1$. From (\ref{Wexp}), it is clear that
the limit $s \to \half$ is regular for any fixed $t$. Also, the limit $t^2 \to
1$ is regular, since both coefficients $\hat a_k(t)$ and $\hat b_k(t)$ have
smooth limits. Thus, in both the crossed s- and u-channels, the function $\bar
W$ is smooth and produces no logarithmic contributions.

We are now ready to sum all contributions in the crossed channels.
For graviton exchange, we have
\bea
I_{\rm grav} ^{(t)} & \sim & - {2^8 \cdot 3^2 \over 7 \pi ^6 (x_{13}^2)^8}
\ln \frac{x_{12}^2 x_{34}^2}{x_{13}^4}
\qquad {\rm s-channel} \qquad
\\
I_{\rm grav} ^{(t)} & \sim & - {2^8 \cdot 3^2 \over 7 \pi ^6 (x_{12}^2)^8}
\ln \frac{x_{14}^2 x_{23}^2}{x_{12}^4}
\qquad {\rm u-channel} \qquad
\eea
For $I_q^{\rm eff}$, we have
\bea
I_q^{\rm eff}
& \sim &
    - {2^6 \cdot 3^3 \over 7 \pi ^6 (x_{13}^2)^8}
    \ln \frac{x_{12}^2 x_{34}^2}{x_{13}^4}
\qquad {\rm s-channel} \qquad
\\
I_q^{\rm eff}
& \sim &
    - {2^6 \cdot 3^3 \over 7 \pi ^6 (x_{12}^2)^8}
    \ln \frac{x_{14}^2 x_{23}^2}{x_{12}^4 }
\qquad {\rm u-channel} \qquad
\eea

We can finally assemble the leading logarithmic asymptotics for the
full amplitudes of normalized operators. For the
4--dilaton and 4--axion amplitudes the limits in the various
channel are equivalent, so we need only quote
\be \label{4dillog}
\langle \hat \ophi \hat \ophi \hat \ophi \hat \ophi \rangle
=\langle \hat \oc \hat \oc \hat \oc \hat \oc \rangle
\sim  -\frac{1}{N^2} \frac{2^5 \cdot 13}{3 \cdot 7 }
\,\frac{1}{x_{12}^{16}}\ln \frac{x_{13}x_{24}}{x_{12}^2}\qquad {\rm
t-channel} \qquad
\ee
The amplitude $\langle \hat \ophi \hat \oc \hat \ophi \hat \oc \rangle$
admits the two different limits (the s-- and u--channels are equivalent):
\bea \label{dilaxt}
\langle \hat \ophi \hat \oc \hat \ophi \hat \oc \rangle
& \sim & \frac{1}{N^2} \frac{2^8 }{3 \cdot 7 }
\,\frac{1}{x_{12}^{16}}\ln \frac{x_{13}x_{24}}{x_{12}^2}\qquad {\rm
t-channel} \qquad \\
 \label{dilaxs}
\langle \hat \ophi \hat \oc \hat \ophi \hat \oc \rangle
&\sim& -\frac{2^4}{N^2} \,\frac{1}{x_{13}^{16}}\ln \frac{x_{12}x_{34}}{x_{13}^2}\quad \qquad {\rm
s-channel} \qquad
\eea

\end{document}